\RequirePackage{lineno} 

\documentclass[twocolumn,showpacs,preprintnumbers,amsmath,amssymb,nofootinbib]{revtex4}

\usepackage{graphicx}
\usepackage{dcolumn}
\usepackage{bm}
 
\usepackage{epstopdf}
\def\bea{\begin{eqnarray}}
\def\eea{\end{eqnarray}}

\def\pp{\mbox{$p$-$p$}}

\def\pa{\mbox{$p$-A}}

\def\auau{\mbox{Au-Au}}

\def\pbpb{\mbox{Pb-Pb}}
\def\ppb{\mbox{$p$-Pb}}
\def\dau{\mbox{$d$-Au}}
\def\pn{\mbox{$p$-N}}
\def\aa{\mbox{A-A}}
\def\nn{\mbox{N-N}}

\def\ee{\mbox{$e^+$-$e^-$}}

\def\pt{$p_t$}
\def\mt{$m_t$}
\def\yt{$y_t$}

\def\nch{$n_{ch}$}
\def\mmpt{$\bar p_t$}

\begin{document} 

\setlength{\pdfpagewidth}{8.5in}
\setlength{\pdfpageheight}{11in}

\setpagewiselinenumbers
\modulolinenumbers[5]

\preprint{version 1.2}

\title{Precision identified-hadron spectrum analysis for 5 TeV $\bf p$-$\bf Pb$ collisions -- Part II
}

\author{Thomas A.\ Trainor}\affiliation{University of Washington, Seattle, WA 98195}


\date{\today}

\begin{abstract}

This is the second part of a two-part article on precision modeling of identified-hadron (PID) $p_t$ spectra from 5 TeV $p$-Pb collisions. In Part I a revised two-component (soft + hard) model (TCM) of PID spectra was introduced, an apparent detection inefficiency for protons was corrected and jet-related spectrum hard components isolated via an improved method were compared to a fixed TCM serving as reference. In the present article (Part II) the TCM is further elaborated to describe centrality variation of spectrum hard components (and therefore entire spectra) within data statistical uncertainties as determined by a standard statistical measure (Z-scores). The completed PID TCM is then used to investigate properties of spectrum and yield ratios (e.g.\ $p/\pi$ ratios) and ensemble-mean $\bar p_t$ data. Systematic differences are observed between meson and baryon hard components. With increasing $p$-Pb centrality meson hard components shift to lower $p_t$ while baryon hard components shift to higher $p_t$. Those trends explain quantitatively the main features and centrality variation of spectrum ratios. Ensemble-mean $\bar p_t$ trends are also predicted at the level of data statistical uncertainties, and details of the $\bar p_t$ centrality trends are explained quantitatively.

\end{abstract}

\pacs{12.38.Qk, 13.87.Fh, 25.75.Ag, 25.75.Bh, 25.75.Ld, 25.75.Nq}

\maketitle

 \section{Introduction}
 
This article presents Part II of a two-part study of identified-hadron (PID) \pt\ spectra from 5 TeV \ppb\ collisions~\cite{aliceppbpid}. In Part I~\cite{pidpart1} a PID spectrum two-component (soft + hard) model (TCM) introduced in Ref.~\cite{ppbpid} is extended in several ways, including incorporation of a resonance contribution to the pion soft-component model. The proton spectra are corrected for an apparent detection inefficiency.  Coefficients $z_{si}(n_s)$ and $z_{hi}(n_s)$ are inferred directly from spectra and modeled by simple functions of TCM hard/soft ratio $x\nu$. Minimum-bias data hard components $H_i(y_t,n_s)$ (for hadron species $i$) are derived from published spectra with a more precise technique. Evolution of data hard components with centrality is determined relative to the fixed PID TCM as a reference. And comparison is made between the revised TCM and the preliminary version described in Ref.~\cite{ppbpid}.
 
In the present study PID TCM hard-component models $\hat H_{0i}(y_t,n_s)$ are modified to accommodate variation of data hard components with charge multiplicity \nch. Systematic differences between meson and baryon trends are revealed. The resulting accuracy of the overall TCM spectrum model is evaluated with the Z-score statistic. The PID TCM is employed to predict spectrum and yield ratios compared with corresponding data. The comparison provides jet-related explanations for critical spectrum ratio features. The PID TCM is employed to predict ensemble-mean \mmpt\ centrality trends. Certain small-amplitude systematic features are explained by variation of jet-related TCM hard components. Several aspects of those results support the validity of a correction for proton detection inefficiency derived in Part I. Some theoretical analysis is also consistent with the possibility of significant proton detection inefficiency~\cite{stoeckerstatmodel,thermalprotons}.

The overall two-part study responds to a tendency in recent years to interpret similarities between certain data features derived from smaller collision systems and corresponding features in \aa\ data as indicating formation of a quark-gluon plasma (QGP) in the smaller systems based on {\em argument from analogy}: {\em Assuming} QGP formation in larger collision systems, and given data similarities between smaller and larger systems, then QGP may be formed in smaller systems. As noted in Part I, because PID spectrum data are conventionally used to buttress such arguments it is imperative to establish the fullest possible understanding of spectrum composition and its response to A-B centrality variation, {\em especially as regards contributions from minimum-bias jets}. That is the principal purpose for developing an extended PID TCM capable of exhaustively characterizing PID \pt\ spectra.

This article is arranged as follows:
Section~\ref{spectrumtcm} introduces the basic PID spectrum TCM developed in Part I and reviews 5 TeV \ppb\ PID spectrum data.
Section~\ref{tcmadapt} describes a newly-developed variable TCM wherein hard-component model functions are adjusted to accommodate spectrum data.
Section~\ref{specratios} presents TCM analysis of PID spectrum and yield ratios with explanations for certain prominent features.
Section~\ref{pidmmpt} describes application of the variable PID TCM to \ppb\ ensemble-mean \mmpt\ data.
Section~\ref{sys}  reviews systematic uncertainties.
Sections~\ref{disc} and~\ref{summ} present discussion and summary.

\section{$\bf p$-$\bf Pb$ PID Spectrum $\bf TCM$} \label{spectrumtcm}

Part I of this study, reported in Ref.~\cite{pidpart1}, addressed several issues: (a) A proton detection inefficiency is quantitatively evaluated and corrected. (b) The centrality dependence of PID TCM coefficients $z_{si}(n_s)$ and $z_{hi}(n_s)$ is obtained directly from PID spectra. (c) That centrality dependence is incorporated into the PID TCM and \pt\ spectra are reanalyzed with greater accuracy. This section presents the basic PID spectrum TCM from Part I including those elements and introduces the PID spectrum data that are the subject of analysis.

\subsection{p-Pb spectrum TCM for identified hadrons} \label{pidspecc}

To establish a TCM for A-B PID \pt\ spectra it is assumed that (a) \nn\ (i.e.\ nucleon-nucleon) parameters $\alpha$, $\bar \rho_{sNN}$ and $\bar \rho_{hNN}$ have been inferred from unidentified-hadron (nonPID) data and (b) geometry parameters $N_{part}$, $N_{bin}$, $\nu$ and $x$ are a common property (relating to centrality) of a specific A-B collision system independent of specific hadron species. Quantity $n_s = \Delta \eta \bar \rho_s$ (referring to the soft component of total charge density $\bar \rho_0 = \bar \rho_s + \bar \rho_h$) is employed as a centrality index. Transverse rapidity  $y_{ti} \equiv \ln[(p_t + m_{ti})/m_i]$ is define for hadron species $i$. For pion rapidity $y_{t\pi}$ the correspondence with \pt\ is 1 vs 0.16 GeV/c, 2 vs 0.5 GeV/c, 2.67 vs 1 GeV/c, 4 vs 3.8 GeV/c and 5 vs 10.4 GeV/c. $y_{t\pi}$ is the default rapidity for all spectrum plots appearing in this study.

Given the A-B spectrum TCM for unidentified-hadron spectra a corresponding TCM for identified hadrons can be defined by assuming that each hadron species $i$ comprises certain {\em fractions} of soft and hard TCM components denoted by $z_{si}(n_s)$ and $z_{hi}(n_s)$ and assumed {independent of \yt}. The PID spectrum TCM is then expressed as
\bea \label{pidspectcm}
\bar \rho_{0i}(y_t) &=& S_i(y_t,n_s)+ H_i(y_t,n_s)
\\ \nonumber
&\approx& z_{si}(n_s)  \bar \rho_{s} \hat S_{0i}(y_t) +  z_{hi}(n_s)  \bar \rho_{h} \hat H_{0i}(y_t) 
\nonumber \\ \label{eq4}
\frac{\bar \rho_{0i}(y_t)}{ z_{si}(n_s)  \bar \rho_{s}} &\equiv&   X_i(y_t)
\\ \nonumber
&\approx & \hat S_{0i}(y_t) +  \tilde z_{i}(n_s)x(n_s)\nu(n_s) \hat H_{0i}(y_t),
\eea
where $\tilde z_{i}(n_s) \equiv z_{hi}(n_s) / z_{si}(n_s)$ and unit-integral model functions $\hat S_{0i}(y_t)$ and $\hat H_{0i}(y_t)$ depend on hadron species $i$. The nonPID factors are defined in terms of \nn\ quantities by $\bar \rho_s = (N_{part}/2) \bar \rho_{sNN}$ and $\bar \rho_h = N_{bin} \bar \rho_{hNN}$ with $\bar \rho_{hNN} \approx \alpha \bar \rho_{sNN}^2$ for linear superposition of \nn\ collisions in A-B collisions and with $\bar \rho_h / \bar \rho_s = x(n_s)\nu(n_s)$. Soft fractions $z_{si}(n_s)$ can be expressed within the TCM by
\bea \label{rhosi}
z_{si}(n_s)&=& \left[\frac{1 + x(n_s) \nu(n_s)}{1 + \tilde z_{i}(n_s) x(n_s) \nu(n_s)} \right]  {z_{0i}},
\eea
with $z_{hi}(n_s) = \tilde z_{i}(n_s) z_{si}(n_s)$
and $\bar \rho_{0i} \equiv z_{0i} \bar \rho_0$ defining $z_{0i}$. Thus, if $\tilde z_{i}(n_s)$ and $z_{0i}$ are specified for relevant hadron species then the remainder of the PID TCM is determined. Model functions $\hat S_{0i}(y_t)$ are defined on proper $m_{ti}$ for a given hadron species $i$ and then transformed to $y_{t\pi}$. $\hat H_{0i}(y_t)$ are always defined on $y_{t\pi}$. The basis for the $\hat S_{0i}(y_t)$ model definitions is the limit of normalized data spectra $X_i(y_t) $ as $n_{ch} \rightarrow 0$. Soft-component models are thus always derived from data spectra. That description relates to a {\em fixed} TCM in which model function $\hat H_{0i}(y_t)$ does not vary with \nch. Elaboration to a {\em varying} TCM that accommodates specific data is described in Sec.~\ref{tcmadapt}.

Normalized data spectra $X_i(y_t)$ can be combined with soft-component model function $\hat S_{0i}(y_t)$ per Eq.~(\ref{eq4}) to extract spectrum data hard components in the form
\bea \label{pidhard}
Y_i(y_t) &\equiv& \frac{1}{\tilde z_{i}(n_s) x(n_s) \nu(n_s) } \left[ X_i(y_t) -  \hat S_{0i}(y_t) \right]~~~
\eea
that may then be compared directly with hard-component model functions $\hat H_{0i}(y_t)$.

\subsection{TCM model parameters}

Table~\ref{rppbdata} presents TCM geometry parameters for 5 TeV \ppb\ collisions inferred from the analysis in Ref.~\cite{tomglauber}. Those geometry parameters, derived from \ppb\ \pt\ spectrum and \mmpt\ data for unidentified hadrons~\cite{alicempt,tommpt,tomglauber}, are assumed to be valid also for each identified-hadron species and were used unchanged to process \ppb\ PID spectrum data in Ref.~\cite{ppbpid}. 
The $\bar \rho_0 = n_{ch} / \Delta \eta$ charge densities are measured quantities inferred from Fig.~16 of Ref.~\cite{aliceglauber}. Specifically, charge density distributions from that figure were averaged over $|\eta_\text{lab}| < 0.5$. Relations $N_{part} = N_{bin} + 1$ and $\nu = 2 N_{bin} / N_{part}$ involve the number of nucleon N participants and \nn\ binary collisions. $\bar \rho_{sNN}$ is the mean soft-component charge density per participant pair averaged over all pairs. $x \equiv \bar \rho_{hNN}/\bar \rho_{sNN} \approx \alpha \bar \rho_{sNN}$ is the hard/soft density ratio. That approximation is based on TCM results from \pp\ \pt\ spectra~\cite{ppprd,ppquad} and is applied here also to \ppb\ collisions assuming they consist of linear superpositions of \nn\ collisions. Columns $\sigma' / \sigma_0$ and $N_{bin}'$ present nominal centralities (bin centers) and binary-collision numbers quoted by Ref.~\cite{aliceppbpid} in connection with measured charge densities $\bar \rho_0$, whereas columns $\sigma / \sigma_0$ and $N_{bin}$ present values inferred in Ref.~\cite{tomglauber} (see Ref.~\cite{tomexclude} for a possible mechanism for such large differences). The remaining table values are a result of the latter analysis.

\begin{table}[h]
	\caption{TCM fractional cross sections $\sigma / \sigma_0$ (bin centers) and  geometry parameters, midrapidity charge density $\bar \rho_0$, \nn\ soft component $\bar \rho_{sNN}$ and TCM hard/soft ratio $x(n_s)$ used for 5 TeV \ppb\ PID spectrum analysis~\cite{tomglauber}. Centrality parameters are from Ref.~\cite{tomglauber}. $\sigma' / \sigma_0$ values are from Table~1 of Ref.~\cite{aliceppbpid}.
	}
	\label{rppbdata}
\begin{center}
	\begin{tabular}{|c|c|c|c|c|c|c|c|c|} \hline
		$n$ &   $\sigma' / \sigma_0$ & $N_{bin}'$  &  $\sigma / \sigma_0$     & $N_{bin}$  & $\nu$ & $\bar \rho_0$ & $\bar \rho_{sNN}$ & $x(n_s)$ \\ \hline
		1	   &      0.025  & 14.7  & 0.15   & 3.20   & 1.52 & 44.6 & 16.6  & 0.188 \\ \hline
		2	 &  0.075  & 13.0 & 0.24    & 2.59   & 1.43 & 35.9 &15.9  & 0.180 \\ \hline
		3	 &  0.15  & 11.7 & 0.37 & 2.16  &  1.37 & 30.0  & 15.2  & 0.172 \\ \hline
		4	 &  0.30 & 9.36 & 0.58  & 1.70   & 1.26  & 23.0  & 14.1  & 0.159  \\ \hline
		5	 &  0.50  & 6.42 &0.80    & 1.31   & 1.13 & 15.8 &   12.1 & 0.137  \\ \hline
		6	 &  0.70 & 3.81 & 0.95   & 1.07   & 1.03  & 9.7  &  8.7 & 0.098 \\ \hline
		7	 & 0.90  & 1.94 & 0.99  & 1.00  & 1.00  &  4.4  & 4.2 &0.047  \\ \hline
	\end{tabular}
\end{center}
\end{table}

Table~\ref{pidparamsxx} presents hard-component model parameters based on optimized descriptions of spectrum hard components for the {\em most-central} ($n = 1$) \ppb\ event class as described in Ref.~\cite{pidpart1}. That condition plus the more-differential format in Part I leads to improved precision of inferred parameter values relative to the preliminary study reported in Ref.~\cite{ppbpid}. TCM soft-component parameters remain unchanged from those reported in Table II of Part I. That table plus Tables~\ref{rppbdata} and \ref{pidparamsxx} reported here define a {\em fixed} PID TCM established as a reference.
 
 \begin{table}[h]
 	\caption{Revised PID TCM hard-component model parameters $(\bar y_t,\sigma_{y_t},q)$ for identified hadrons from 5 TeV \ppb\ collisions derived from the differential spectrum analysis in Rev.~\cite{pidpart1}.  Parameter $z_{0i}$ values, inferred as limiting values of $z_{si}(n_s)$ centrality trends, are also included. Uncertainties are determined as one half the change that would produce an obvious variation in $z_{hi}(y_t,n_s)$ ratios. Values with no uncertainties are duplicated from a related particle type.
 	}
 	\label{pidparamsxx}
 	\begin{center}
 		\begin{tabular}{|c|c|c|c|c|} \hline
 			&  $\bar y_t$ & $\sigma_{y_t}$ & $q$ & 	$z_{0i}$  \\ \hline
 			$ \pi^\pm $     
 			&	   $2.46\pm0.005$ & $0.575\pm0.005$ & $4.1\pm0.5$  &  0.82 $\pm$0.01 \\ \hline
 			$K^\pm$    
 			&	  $2.655$  & $0.568$ & $4.1$  & 0.128 $\pm$0.002 \\ \hline
 			$K_\text{S}^0$          
 			& 	   $2.655\pm0.005$ & $0.568\pm0.003$ & $4.1\pm0.1$ & 0.064 $\pm$0.002  \\ \hline
 			$p$        
 			& 	  $2.99\pm0.005$  & $0.47\pm0.005$ & $5.0$ & 0.065 $\pm$0.002  \\ \hline
 			$\Lambda$       
 			& 	  $2.99\pm0.005$  & $0.47\pm0.005$ & $5.0\pm0.05$  & 0.034 $\pm$0.002 \\ \hline	
 		\end{tabular}
 	\end{center}
 \end{table}
 
 \subsection{5 TeV $\bf p$-Pb PID  spectrum data} \label{alicedata}
 
 The identified-hadron spectrum data reported in Ref.~\cite{aliceppbpid} and adopted for the present analysis were produced by the ALICE collaboration at the LHC.  The event sample for charged hadrons is 12.5 million non-single-diffractive (NSD) collisions and for neutral hadrons 25 million NSD collisions. Collision events were divided into seven charge-multiplicity \nch\ or \ppb\ centrality classes based on yields in a VZERO-A counter subtending $2.8 < \eta_{lab} < 5.1$ in the Pb direction.
 Hadron species include charged pions $\pi^\pm$, charged kaons $K^\pm$, neutral kaons $K^0_\text{S}$, protons $p,~\bar p$ and Lambdas $\Lambda,~\bar \Lambda$. Spectra for charged vs neutral kaons and particles vs antiparticles are reported to be statistically equivalent. Note that a later presentation of PID spectrum data from the same collision system~\cite{alicepidnew} extends the \pt\ acceptance to 20 GeV/c for pions, charged kaons and protons, but the $p/\pi$ ratio results in its Fig.~10 are essentially the same as in Fig.~2 of Ref.~\cite{aliceppbpid} suggesting that the proton inefficiency is equivalent.
 
\begin{figure}[h]
 	\includegraphics[width=1.65in]{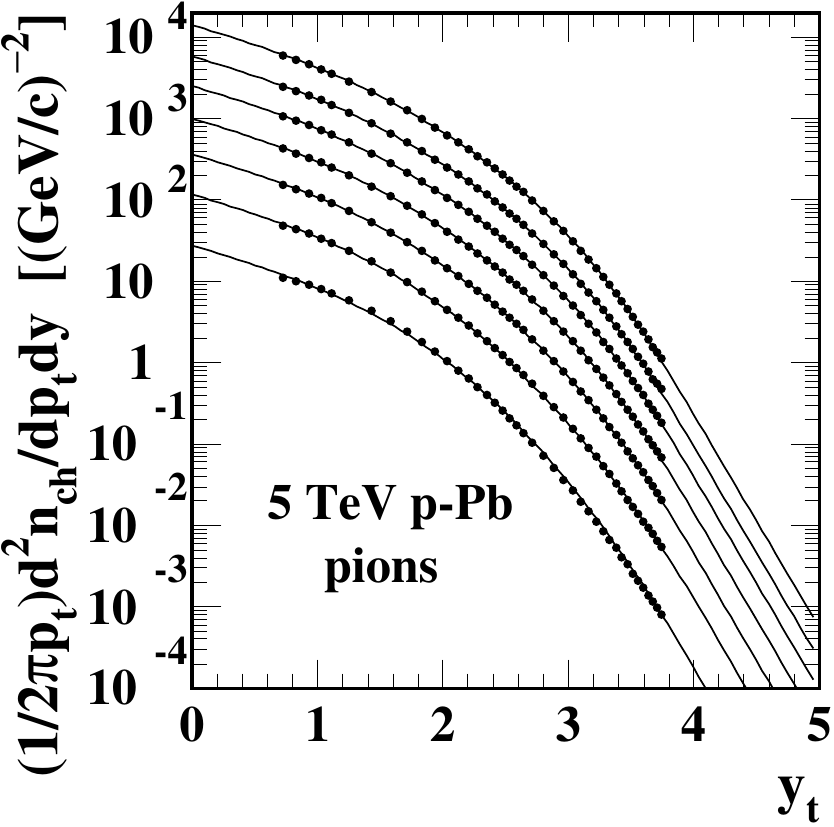}
 	\includegraphics[width=1.65in]{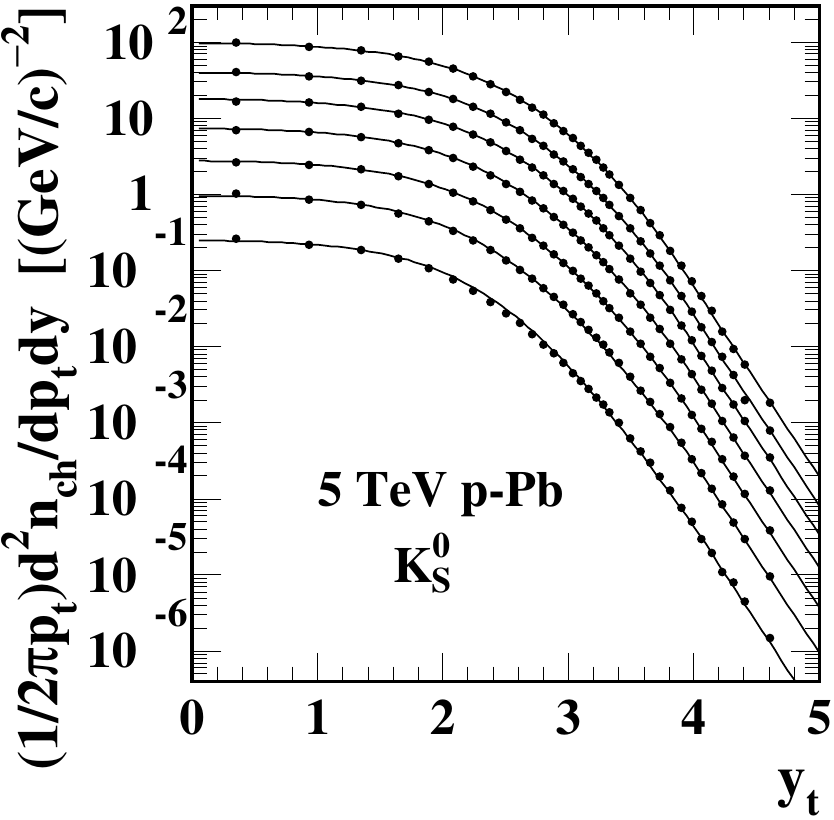}
 	\put(-142,105) {\bf (a)}
 	\put(-23,105) {\bf (b)}\\
 	\includegraphics[width=1.65in]{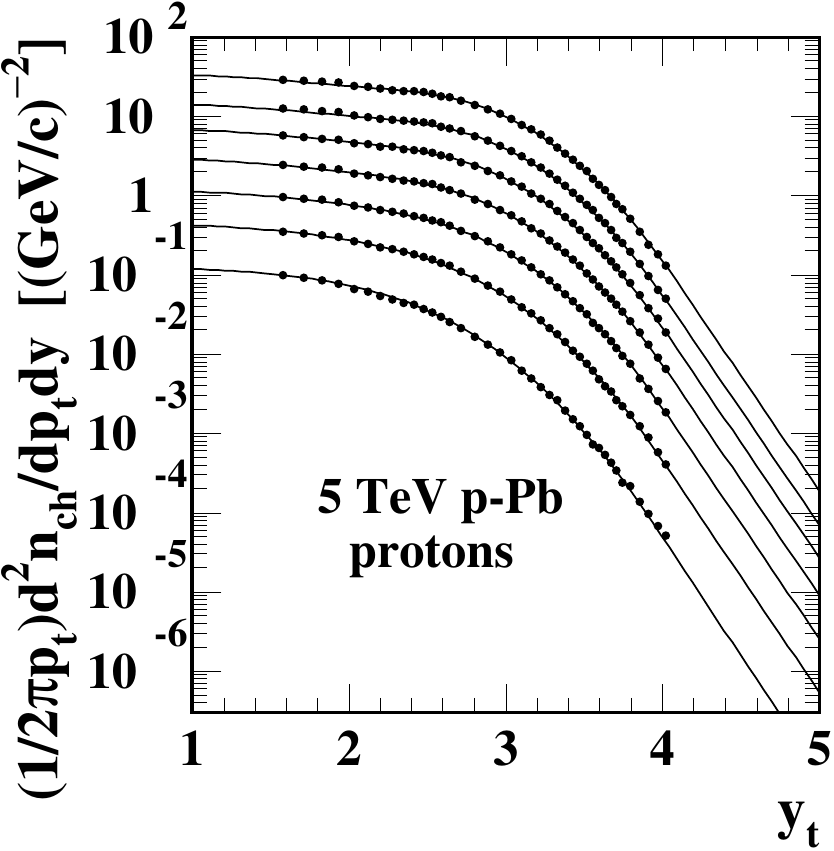}
 	\includegraphics[width=1.65in]{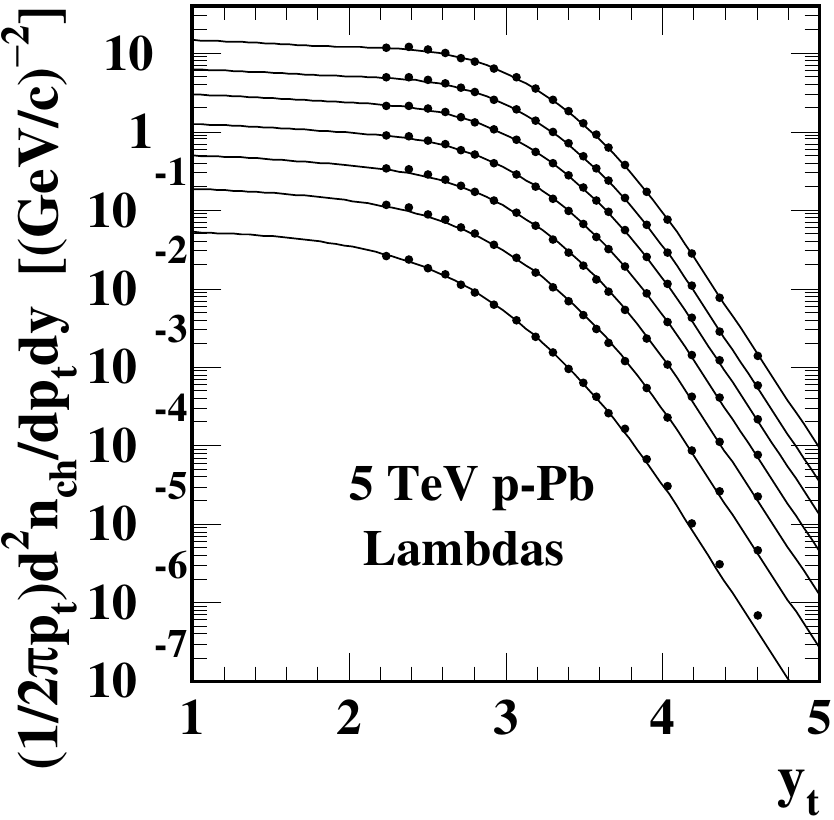}
 	\put(-142,105) {\bf (c)}
 	\put(-23,105) {\bf (d)}\\
 	\caption{\label{piddata}
 		\pt\ spectra for identified hadrons from 5 TeV \ppb~\cite{aliceppbpid} plotted vs pion transverse rapidity \yt\ (default) for:
 		(a) pions,
 		(b) neutral kaons,
 		(c) corrected protons,
 		(d) Lambdas.
 		Solid curves represent the PID spectrum TCM from Sec.~\ref{tcmadapt}.
 	}  
 \end{figure}

 Figure~\ref{piddata} shows PID spectrum data (densities on \pt) from Ref.~\cite{aliceppbpid} (points) plotted vs logarithmic variable \yt. That format (log-log with respect to \pt) provides detailed access to low-\pt\ structure (where most jet fragments appear) and clearly shows power-law trends at higher \pt. The curves are TCM parametrizations derived in Sec.~\ref{tcmadapt} below and demonstrated there  to describe spectrum data within their statistical uncertainties.  As in Ref.~\cite{aliceppbpid} the spectra are scaled up by powers of 2 according to $2^{n-1}$ where $n \in [1,7]$ is the centrality class index and $n = 7$ is most central (following the usage in Ref.~\cite{aliceppbpid}). In this article $n=7$ denotes the {\em least}-central data as in Table~\ref{rppbdata}. Even with the improvement of introducing logarithmic independent variable \yt\ this method of presenting data provides no significant information on the substantial variation of spectrum structure with \ppb\ centrality.
 
\section{Variable TCM adapted to data} \label{tcmadapt}

In this section the precision PID data spectrum hard components isolated in Part I are described accurately with a {\em variable} TCM. Factorization of data hard components $H_i(y_t,n_s)$ is modified to minimize bias. Centrality-dependent model functions $\hat H_0(y_t,n_s)$ are inferred from the minimally-biased data hard components. The revised model parametrization presents the opportunity to interpret modification of jet-related hard components in terms of physical mechanisms. The variable TCM appears to exhaust all information carried by \ppb\ PID \pt\ spectra.

\subsection{TCM hard-component algebra}

In Part I of this study, as reported in Ref.~\cite{pidpart1}, PID spectrum hard components were reported in the form
\bea \label{zhizz}
Y'_i(y_t,n_s)&=&\frac{\bar \rho_{0i}(y_t,n_s) -  z_{si}(n_s)\bar \rho_s \hat S_{0i}(y_t)}{\bar \rho_h  \hat H_{0i}(\bar  y_t)}
\\ \nonumber 
&\equiv& z_{hi}(y_t,n_s) \frac{\hat H_{0i}(y_t)}{\hat H_{0i}(\bar  y_t)}
\eea
where $\bar  y_t$ is in this case the mode of the $\hat H_{0i}(y_t)$ model function, in which case $Y'_i(\bar y_t,n_s) \approx z_{hi}(\bar y_t,n_s)$.	An advantage of the first line is that spectrum hard components are isolated with minimum bias, being scaled only by universal centrality parameter $\bar \rho_h = N_{bin} \alpha \bar \rho_{sNN}^2$ (parameters from Table~\ref{rppbdata}) and fixed amplitude $\hat H_{0i}(\bar  y_t)$. The second line can be solved for $z_{hi}(y_t,n_s)$ to reveal deviations of spectrum data from fixed model function $\hat H_{0i}(y_t)$. Those results were reported in Sec.~VI of Part I.

In the present study the intent is to vary TCM model functions so as to accommodate details of PID hard-component evolution with \ppb\ centrality. As noted, the hard component of spectrum data is best estimated via
\bea
H_i(y_t,n_s)  &\approx& \bar \rho_{0i}(y_t,n_s) -  z_{si}(n_s)\bar \rho_s \hat S_{0i}(y_t),
\eea
where $z_{si}(n_s)$ is determined as accurately as possible, individually for each centrality, at low \yt\ where $H_i(y_t,n_s)$ is negligible. $z_{si}(n_s)\bar \rho_s \hat S_{0i}(y_t)$ is then defined as the asymptotic limit of $\bar \rho_{0i}(y_t,n_s)$ as $n_s \rightarrow 0$ so as to determine $\hat S_{0i}(y_t)$.  That differential procedure is most accurate for high-mass hadrons where the hard/soft ratio is greatest (see Fig.~8 of Part I). The question then becomes how to best factorize $H_i(y_t,n_s)$, a peaked distribution with possibly variable centroid and shape, to characterize it.

\vskip .1in

The initial factorization strategy, defined in Eq.~(1) of Ref.~\cite{ppbpid}, was $H_i(y_t,n_s) \approx z_{hi}(n_s) \bar \rho_h(n_s) \hat H_{0i}(y_t)$ with $\hat H_{0i}(y_t)$ a fixed model function, the default for the TCM. That factorization is insufficient because the shapes of data $H_i(y_t,n_s)$ are observed to change significantly with \ppb\ centrality. To accommodate hard-component shape changes two alternative factorizations are possible:
\bea
H_i(y_t,n_s)&\approx& z_{hi}(y_t,n_s) \bar \rho_h \hat H_{0i}(y_t)
\\ \nonumber
&\approx& z_{hi}(\bar y_t,n_s) \bar \rho_h \hat H_{0i}(y_t,n_s),
\eea
representing centrality-dependent \yt\ shape variation by $z_{hi}(y_t,n_s)$ (first line) or $\hat H_{0i}(y_t,n_s)$ (second line). The first line includes fixed idealized model function $\hat H_{0i}(y_t)$ defined in terms of a specific \ppb\ centrality (e.g.\ $n = 1$ or most central). That strategy is most effective for displaying data-model deviations as in Sec.~VI of Part I but not for adapting the TCM to accommodate changing data structure. The second line includes variable model function $\hat H_{0i}(y_t,n_s)$ that {\em is} adjusted to accommodate data, with $z_{hi}(\bar y_t,n_s)$ estimated at the {\em data} mode which may be centrality dependent. That strategy is most effective for understanding what aspects of the hard component change with centrality and how they may be interpreted physically. The first strategy was employed in Part I as noted in its Sec.~VI. Results from the second strategy are reported below.
Spectrum hard-component variations with \ppb\ centrality appear to be substantially different for mesons (pions, kaons) and baryons (protons, Lambdas). The two cases are considered separately.

\subsection{Meson hard-component models}

The hard components for mesons are observed to have peak modes approximately independent of \ppb\ centrality within data uncertainties. However, the peak widths are observed to vary with centrality above {\em and/or} below the mode. The default TCM model function $\hat H_{0}(y_t)$ consists of a Gaussian on \yt\ with exponential tail~\cite{hardspec}:
\bea \label{h00x}
\hat H_{0}(y_t) &\approx & A \exp\left\{ - \frac{(y_t - \bar y_t)^2}{2 \sigma^2_{y_t}}\right\}~~~\text{near mode $\bar y_t$}
\\ \nonumber
&\propto &  \exp(- q y_t)~~~\text{for larger $y_t$ -- the tail}.
\eea
The model can be modified to accommodate meson hard components by defining separate widths $\sigma_{y_t+} \rightarrow \sigma_+$ above the mode and $\sigma_{y_t-} \rightarrow \sigma_-$ below the mode. Model function $\hat H_{0i}(y_t,n_s)$ for a given system is then a composite of functions defined below and above the mode.

\begin{figure}[h]
	\includegraphics[width=3.3in]{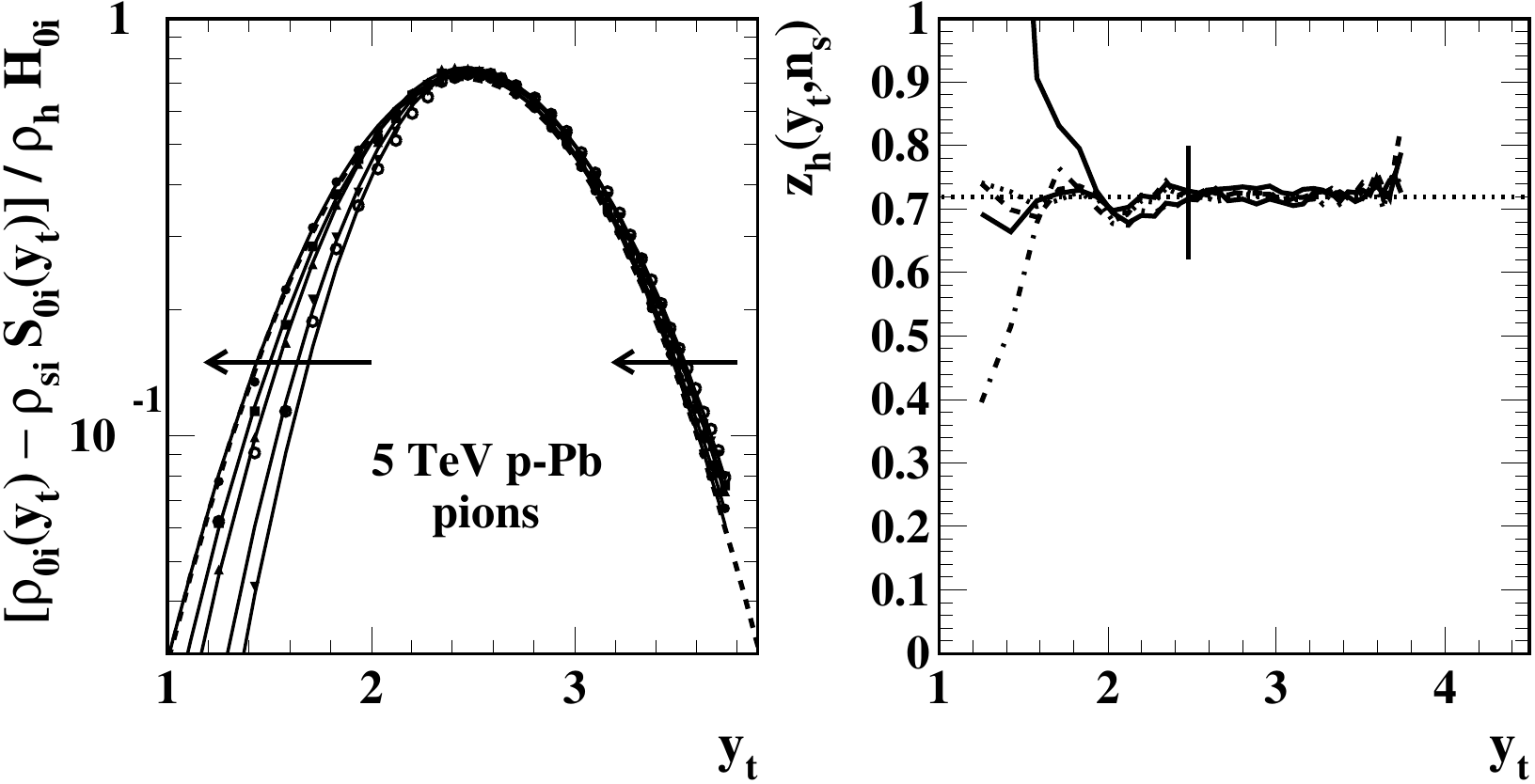}
	\caption{\label{newpion}
		Left: Pion data spectrum hard components (points) for 5 TeV \ppb\ collisions and centralities denoted by $n \in [1,5]$ (see Table~\ref{rppbdata}) in the form of Eq.~(\ref{zhizz}) (first line). The curves are TCM model function $\hat H_{0i}(y_t,n_s)$ plotted in the form appearing on the right-hand side of Eq.~(\ref{pidhcnew}). The arrows indicate effective data shifts with increasing \ppb\ centrality.
		Right: 	Data/model ratio $H_i(y_t,n_s) / \bar \rho_h \hat H_{0i}(y_t,n_s)$ per Eq.~(\ref{pidhcnew}). The dotted line is reference $z_{hi}(\bar y_t,n_s) = 0.72$ for $n = 1$.
	}  
\end{figure}

Figure~\ref{newpion} (left) shows pion hard components in the form 
\bea \label{pidhcnew}
\frac{H_i(y_t,n_s)}{\bar \rho_h \hat H_{0i}(\bar y_t,n_s)} &\approx& z_{hi}(\bar y_t,n_s)\frac{\hat H_{0i}(y_t,n_s)}{\hat H_{0i}(\bar y_t,n_s)},
\eea
where the left-hand side represents data (points) and the right-hand side represents the model (curves). The bold dashed curve is based on $\hat H_{0i}(y_t,n_s)$ for most-central data ($n = 1$) as in Fig.~11 (left) of Part I to provide a reference. Data for the three most-central event classes follow the model down to \yt\ = 1.2 ($p_t \approx 0.20$ GeV/c).

Figure~\ref{newpion} (right) shows $z_{hi}(y_t,n_s)$ 
derived from data via Eq.~(\ref{pidhcnew}) as ratio  $H_i(y_t,n_s)/\bar \rho_h \hat H_{0i}(y_t,n_s)$  (curves) vs the nominal $z_{hi}(\bar y_t,n_s)$ for the most-central data (dotted line). The curves in the right panel thus represent deviations via ratio of data compared to the variable TCM.

\begin{figure}[h]
	\includegraphics[width=3.3in]{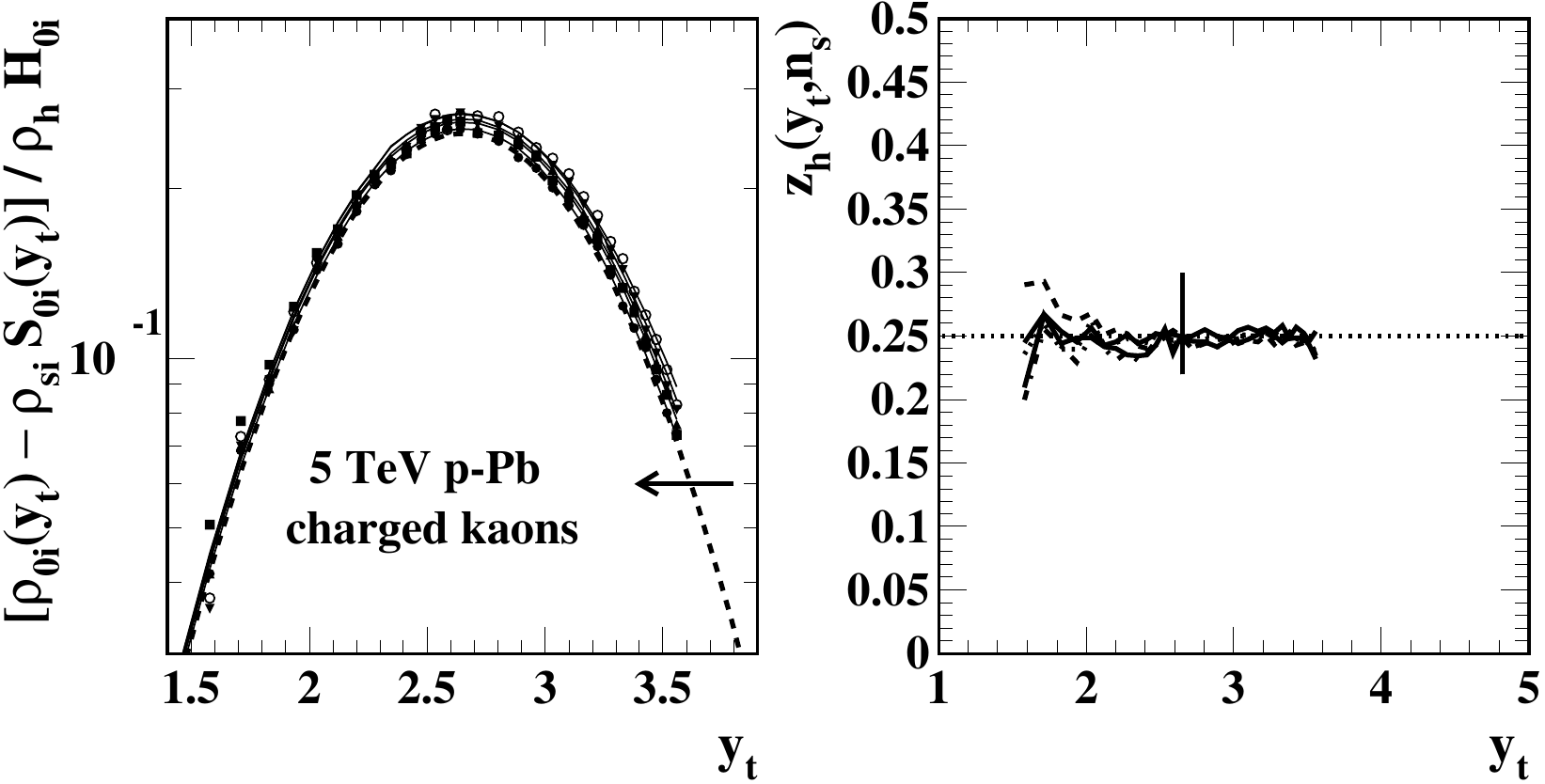}
\put(-202,105) {\bf (a)}
\put(-23,105) {\bf (b)}\\
	\includegraphics[width=3.3in]{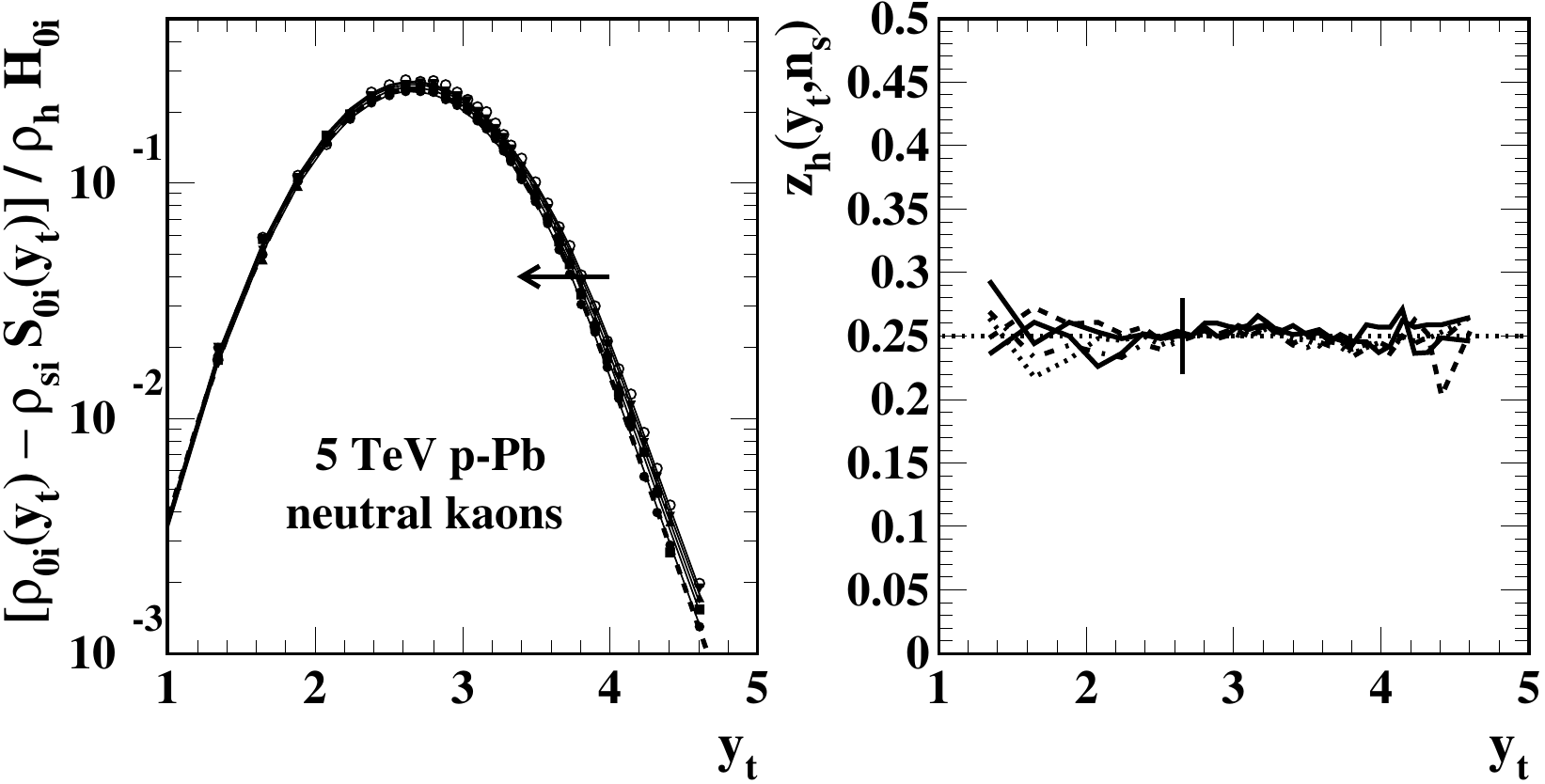}
\put(-202,105) {\bf (c)}
\put(-23,105) {\bf (d)}\\
	\caption{\label{newkaon}
Spectrum data hard components for charged kaons (a,b) and neutral kaons (c,d) processed the same as for pion data in Fig.~\ref{newpion}
}  
\end{figure}

Figure~\ref{newkaon} shows the same procedure applied to kaons. The \yt\ acceptance for charged kaons is rather limited. However, the acceptance for neutral kaons demonstrates that the variable TCM for kaons describes data within statistical uncertainties from 0.2 GeV/c to 7 GeV/c as is evident from panel (d). The axis limits in the left panels have been adjusted in each case to provide maximum sensitivity to data structure in relation to models.

\begin{figure}[h]
	\includegraphics[width=3.3in]{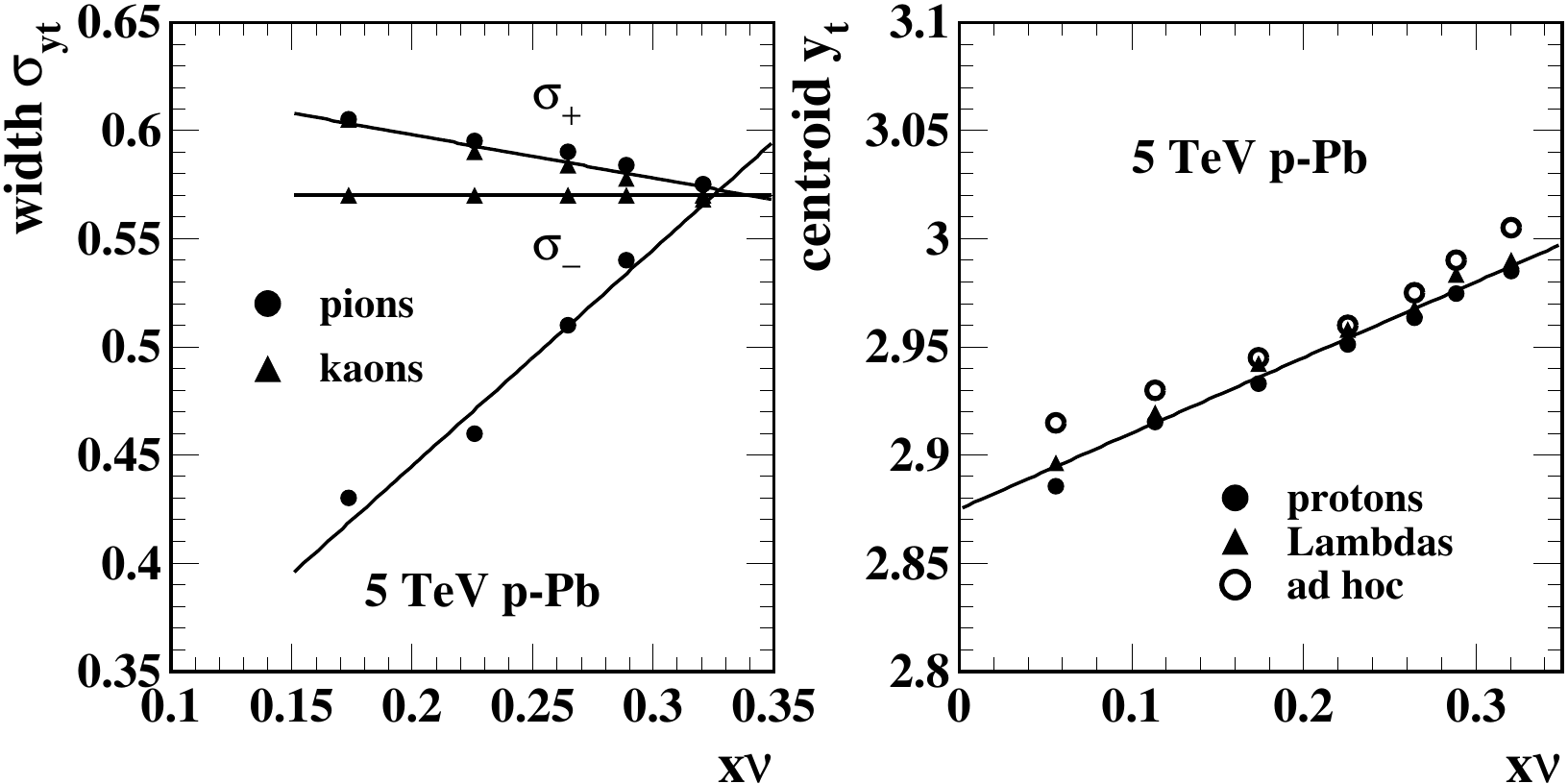}
	\caption{\label{widvar}
	Left: Gaussian widths for TCM hard-component model $\hat H_0(y_t,n_s)$ defined in Eq.~(\ref{h00x}) above the mode $\sigma_+$ and below the mode $\sigma_-$ for pions and kaons.
	Right: 	Hard-component model centroids $\bar y_t$ for protons and Lambdas shifting with \ppb\ centrality according to Eq.~(\ref{barytns}).
} 
\end{figure}

Figure~\ref{widvar} (left) shows $\hat H_{0i}(y_t,n_s)$ model widths above the mode $\sigma_+$ and below the mode $\sigma_-$ vs centrality measure $x\nu$.  The inferred parameter values are consistent with linear trends on $x\nu$ (lines):  $\sigma_+$ for pions and kaons $\approx 0.57 - 0.2(x\nu - 0.34)$ and $\sigma_-$ for pions $\approx 0.57 + (x \nu - 0.325)$. The width below the mode for kaons is held fixed at the value 0.57 obtained for most-central kaon data.

\subsection{Baryon hard-component models}

Peak modes for baryon hard components are observed to shift substantially with \ppb\ centrality, and peak amplitudes [described by $z_{hi}(n_s)$] also vary substantially with centrality. But shifts and amplitudes are correlated such that data hard components coincide with a fixed power-law trend at higher \yt. TCM model function $\hat H_{0i}(y_t,n_s;\bar y_t)$ can accommodate those characteristics if peak mode $\bar y_t$ shifts with amplitude $z_{hi}(n_s)$ according to
\bea  \label{barytns}
\bar y_t(n_s) &=& \bar y_{t0} - (1/q) \ln[z_{hi}(n_s) / z_{hi0}]. 
\eea
following Eq.~(\ref{h00x}) (second line), where $\bar y_{t0}$ and $z_{hi0}$ correspond to a reference centrality (e.g.\ most-central data).
Hard-component data should then be described by
\bea \label{pidhcnewx}
\frac{H_i(y_t,n_s)}{\bar \rho_h \hat H_{0i}(\bar y_t,n_s)} &\approx& z_{hi}(\bar y_t,n_s)\frac{\hat H_{0i}[y_t;\bar y_t(n_s)]}{\hat H_{0i}(\bar y_t,n_s)}
\eea
where the {\em shape} of model $\hat H_{0i}[y_t;\bar y_t(n_s)]$ remains fixed.

\begin{figure}[h]
	\includegraphics[width=3.3in]{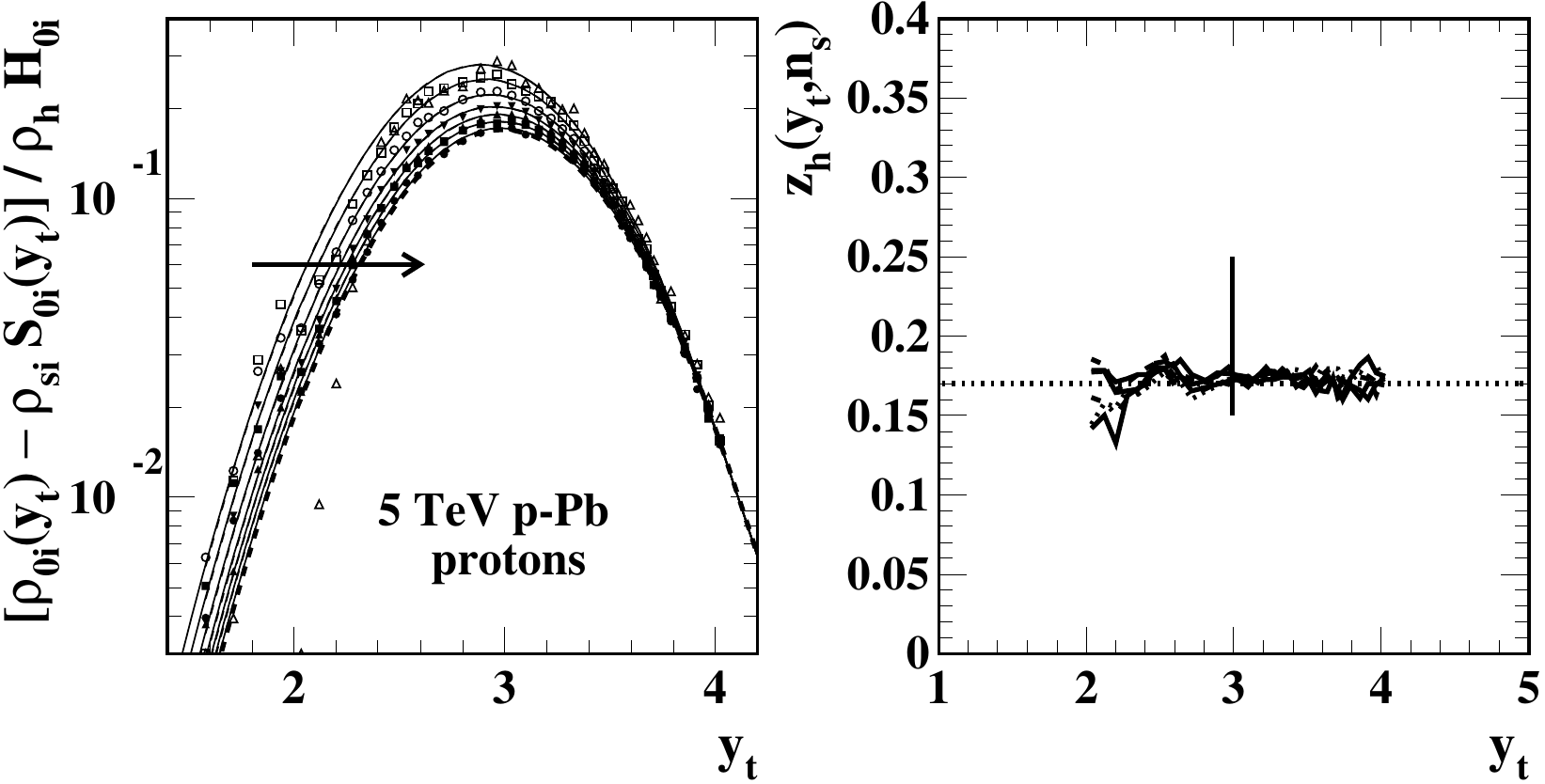}
\put(-202,105) {\bf (a)}
\put(-23,105) {\bf (b)}\\
	\includegraphics[width=3.3in]{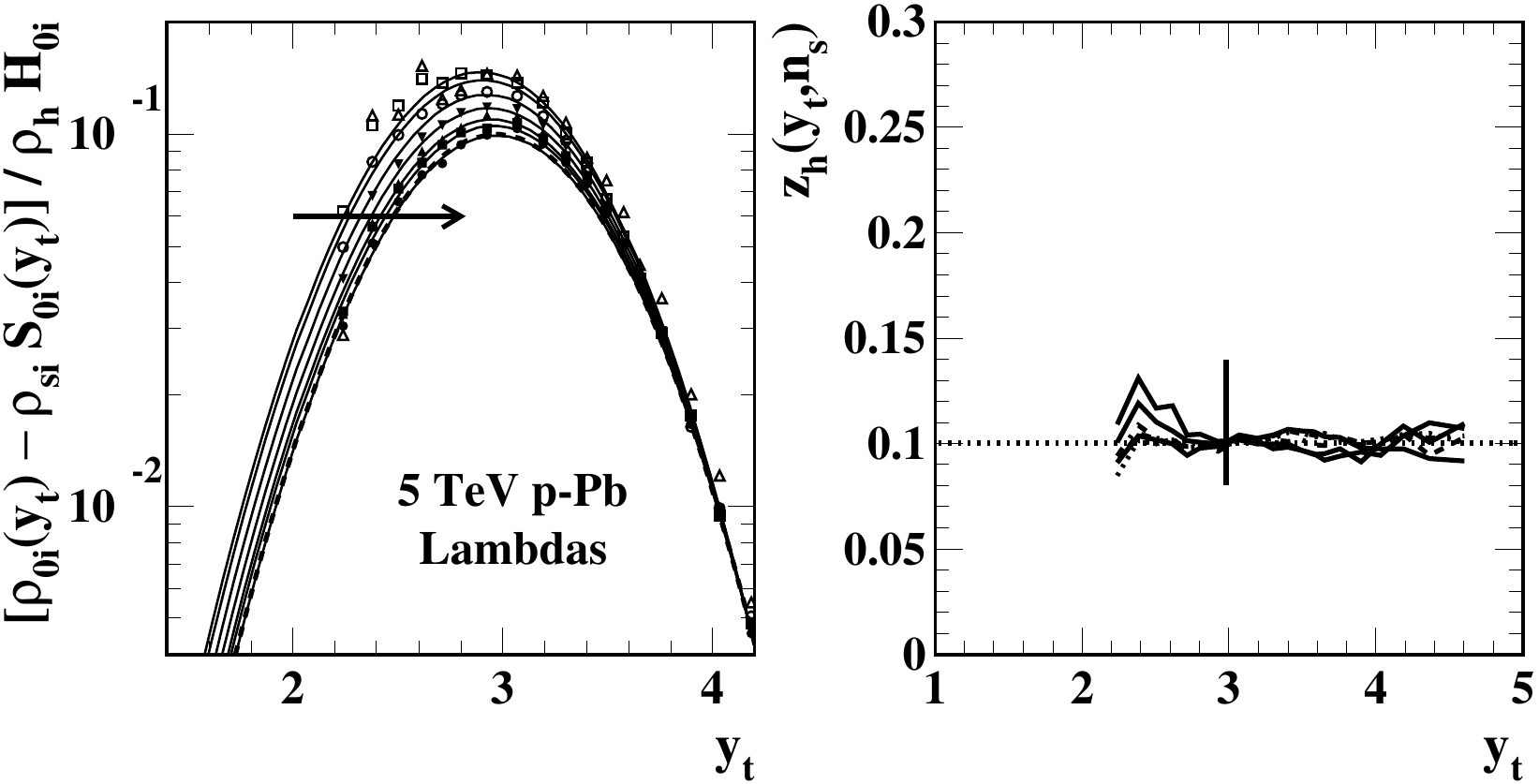}
\put(-202,105) {\bf (c)}
\put(-23,105) {\bf (d)}\\
	\caption{\label{newprot}
Spectrum data hard components for protons (a,b) and Lambdas (c,d) processed the same as pion data in Fig.~\ref{newpion}.
	}  
\end{figure}

Figure~\ref{newprot} shows data-model comparisons for protons (a,b) and Lambdas (c,d). Proton data for more-central collisions follow the model down to \yt\ = 2 ($p_t \approx 0.5$ GeV/c). The quality of the description is evident from the right panels.

Figure~\ref{widvar} (right) shows centroid variation $\bar y_t(n_s)$ for protons and Lambdas (solid points), as determined by $z_{hi}(n_s)$ values from Table V of Part I via Eq.~(\ref{barytns}), vs hard/soft ratio $x\nu$. In Ref.~\cite{ppbpid} the apparent centroid shifts for baryon hard components were modeled by an {\em ad hoc} function
\bea \label{adhoc}
\bar y_t(n_s) &=& 2.96 - \delta_0(n - 4),
\eea
where $n \in [1,7]$ is the centrality index with $n = 1$ most central. The estimated value $\delta_0 = 0.03$ for uncorrected proton spectra was severely biased. The value 0.015 for Lambdas corresponds to the open points in the right panel of Fig.~\ref{widvar} showing reasonable agreement with the present trends (solid points) derived from precise differential measurements. Those inferred values for $\bar y_t(n_s)$ follow a linear trend on centrality measure $x\nu$ (line) 
\bea
\bar y_t(n_s) &=& 2.875 + 0.35 x\nu.
\eea
It is notable that the several parameters controlling variable-TCM PID model functions all vary linearly with hard/soft (jet/nonjet) ratio $x\nu$ within uncertainties.

\subsection{Interpreting hard-component evolution}

Determination of the \nch\ evolution of \ppb\ meson and baryon spectrum hard-component shapes within their point-to-point uncertainties is reported above. A physical interpretation may be attempted based on current understanding of QCD in high-energy nuclear collisions.

The quantitative relation between spectrum hard components and measured jet properties has been established previously~\cite{fragevo}. A spectrum hard component (jet fragment distribution) can be expressed as the convolution of a {\em measured} parton/jet energy spectrum with a {\em measured} ensemble of fragmentation functions (FFs) corresponding to the relevant collision system (e.g.\ \pp\ collisions~\cite{mbdijets} as opposed to \ee\ collisions~\cite{eeprd}). A key parameter is the effective lower bound of the jet spectrum -- near 3 GeV according to NSD \pp\ spectrum data~\cite{mbdijets}.

Hard-component properties are observed to vary with collision charge multiplicity \nch\ that may or may not relate to transverse geometry (centrality) variation. As noted, the \nch\ (i.e.\ $\bar \rho_0$) variation of 5 TeV \ppb\ collisions appearing in Table~\ref{rppbdata} corresponds to rather weak \ppb\ transverse geometry variation. Likewise, the quadratic relation between hard and soft components observed for \pp\ collisions~\cite{ppprd} suggests that \pp\ transverse geometry is not relevant for \pp\ dijet production. Yet for \nch\ variation appearing in Table~\ref{rppbdata} dijet production increases {\em 50 fold} [$\propto N_{bin} \bar \rho_{sNN}^2 \rightarrow (3.2/1) \times (16.6/4.2)^2$]. Such a large increase can only be attributed to the depth of the parton splitting cascade on momentum fraction $x$ corresponding to {\em event-wise} parton distribution functions or PDFs.

Since the combination of PID FF properties and parton energy spectrum determines PID hard components, and the parton energy spectrum is determined in part by event-wise PDFs, one may speculate as to the relation of PID spectrum hard components to event-wise charge density $\bar \rho_0$. In particular, the {\em effective} lower bound of the parton energy spectrum may depend directly on $\bar \rho_0$. What matters is how the parton energy spectrum manifests as jet fragments of a particular hadron species (i.e.\ what is the density of identified-hadron final states).

Evolution of baryon hard components with \nch\ as in Fig.~\ref{newprot} (left) closely resembles that expected for variation of the parton spectrum lower bound as illustrated in Fig.~11 (left) of Ref.~\cite{fragevo} for 200 GeV \pp\ collisions. The solid curve corresponds to 3 GeV and the  neighboring dotted curves correspond to $\pm$ 10\% energy shifts. The net effect for baryons is a lower-bound shift {\em upward} on \yt\ with increasing \nch. In contrast, meson hard-component widths above and/or below the mode respond to increasing \nch\ with the peak centroid shifting {\em downward} on \yt, the net effect being complementary to baryon evolution.  

The relation to a lower bound on the parton energy spectrum appears contradictory for mesons vs baryons. However, what should matter is what partons fragment to what hadron species. Demanding lower-mass mesons may select for lower-energy partons and the converse for higher-mass baryons. It is notable that hard-component evolution trends exhibit remarkably smooth variation vs hard/soft ratio $x\nu$ over an \nch\ interval corresponding to fifty-fold increase in jet production. Over much of that interval \ppb\ collisions are indistinguishable from isolated \pn\ collisions. One may conclude that \ppb\ geometric centrality is not relevant to \pn\ jet modification, a conclusion consistent with an analysis of eventwise-reconstructed jets from 5 TeV \ppb\ collisions~\cite{ppbjets}.

\subsection{Evaluating model accuracy} \label{modelacc}

As noted in Ref.~\cite{tommodeltests} spectrum data-model comparisons are often represented by data/model ratios. Ratio values near 1 are interpreted to indicate acceptable models. However, that procedure can be quite misleading as discussed for instance in Ref.~\cite{ppprd}. A more meaningful measure of model validity is the Z-score~\cite{zscore} defined by
\bea \label{zscore}
Z_i &=& \frac{O_i - E_i}{\sigma_i} \rightarrow \frac{\text{data $-$ model}}{\text{statistical error}},
\eea
where $O_i$ is a spectrum datum, $E_i$ is the corresponding expectation (model prediction) and $\sigma_i$ is the data r.m.s. statistical uncertainty (error). For Poisson-distributed samples $\sigma_i \approx \sqrt{E_i}$ in Eq.~(\ref{zscore}), leading to the formula often encountered for the $\chi^2$ statistic. Given the Z-score definition in Eq.~(\ref{zscore}) the relation to the $\chi^2$ statistic can be simply expressed as
\bea \label{chinu}
\chi^2 & \equiv & \sum_{i=1}^N \frac{(O_i - E_i)^2}{\sigma_i^2} =  \sum_{i=1}^N Z_i^2
\eea
for $N$ data points in a spectrum. Given model degrees of freedom $\nu = N - \text{number of fit parameters}$ one expects $\chi^2 \sim \nu$, in which case the r.m.s. value for Z scores for an acceptable fit should be $\sqrt{\nu / N}$ -- somewhat less than 1.  

The relation between data/model ratios and corresponding Z-scores is
\bea \label{suppress}
\frac{\text{data}}{\text{model}} - 1 &\approx& \text{Z-score} \times \frac{\text{error}}{\text{data}},
\eea
with error/model (exact) $\rightarrow $ error/data (approximate). The error/data ratio (typically $\ll 1$) can vary by orders of magnitude between different particle types and collisions systems, and even across \yt\ intervals. Whereas the correct test of model validity is the relation of Z-scores to 1 the interpretation of data/model ratios relative to 1 is highly problematic and typically quite misleading.

\begin{figure}[h]
	\includegraphics[width=3.3in]{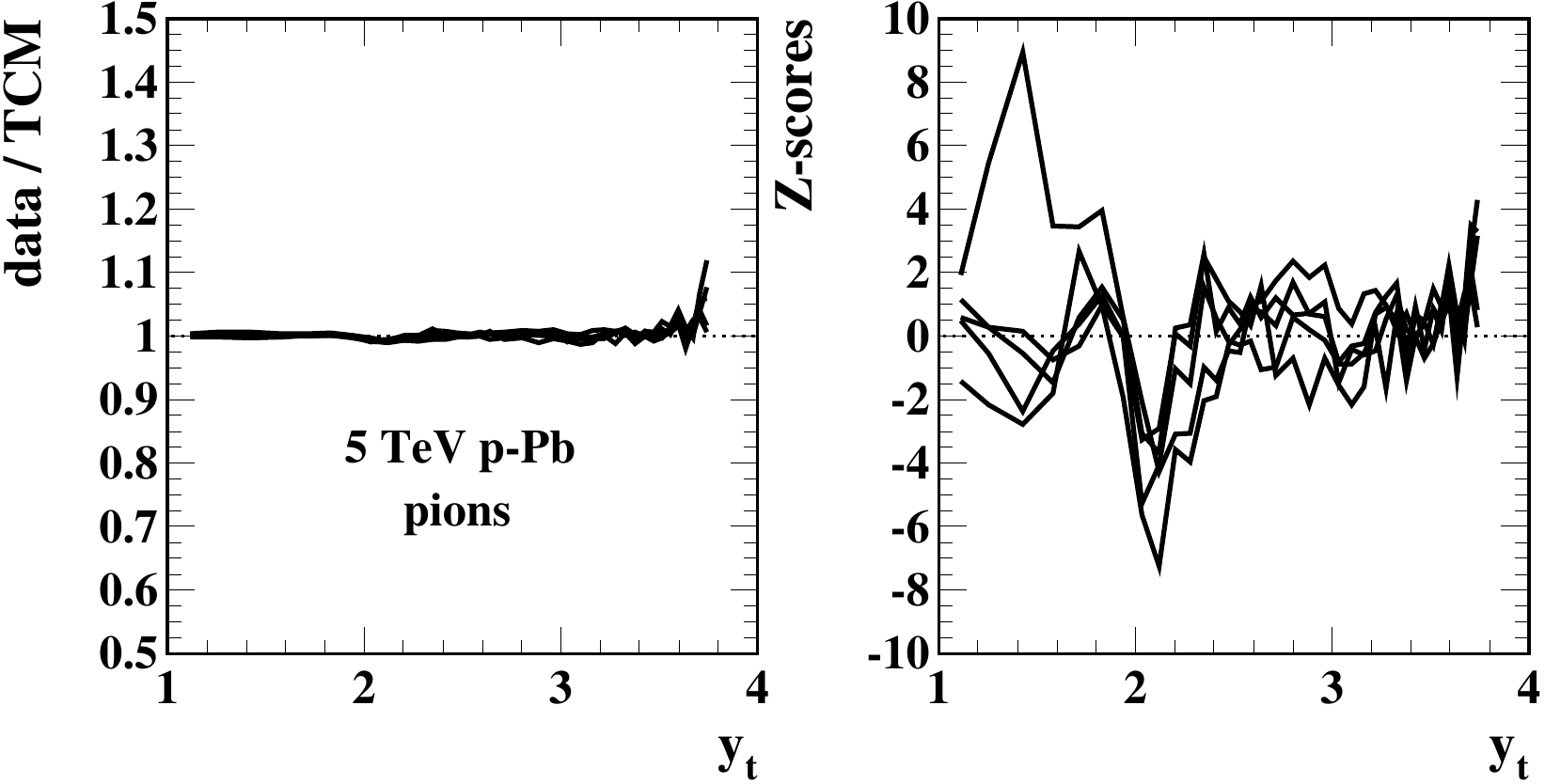}
	\put(-202,105) {\bf (a)}
	\put(-23,105) {\bf (b)}\\
	\includegraphics[width=3.3in]{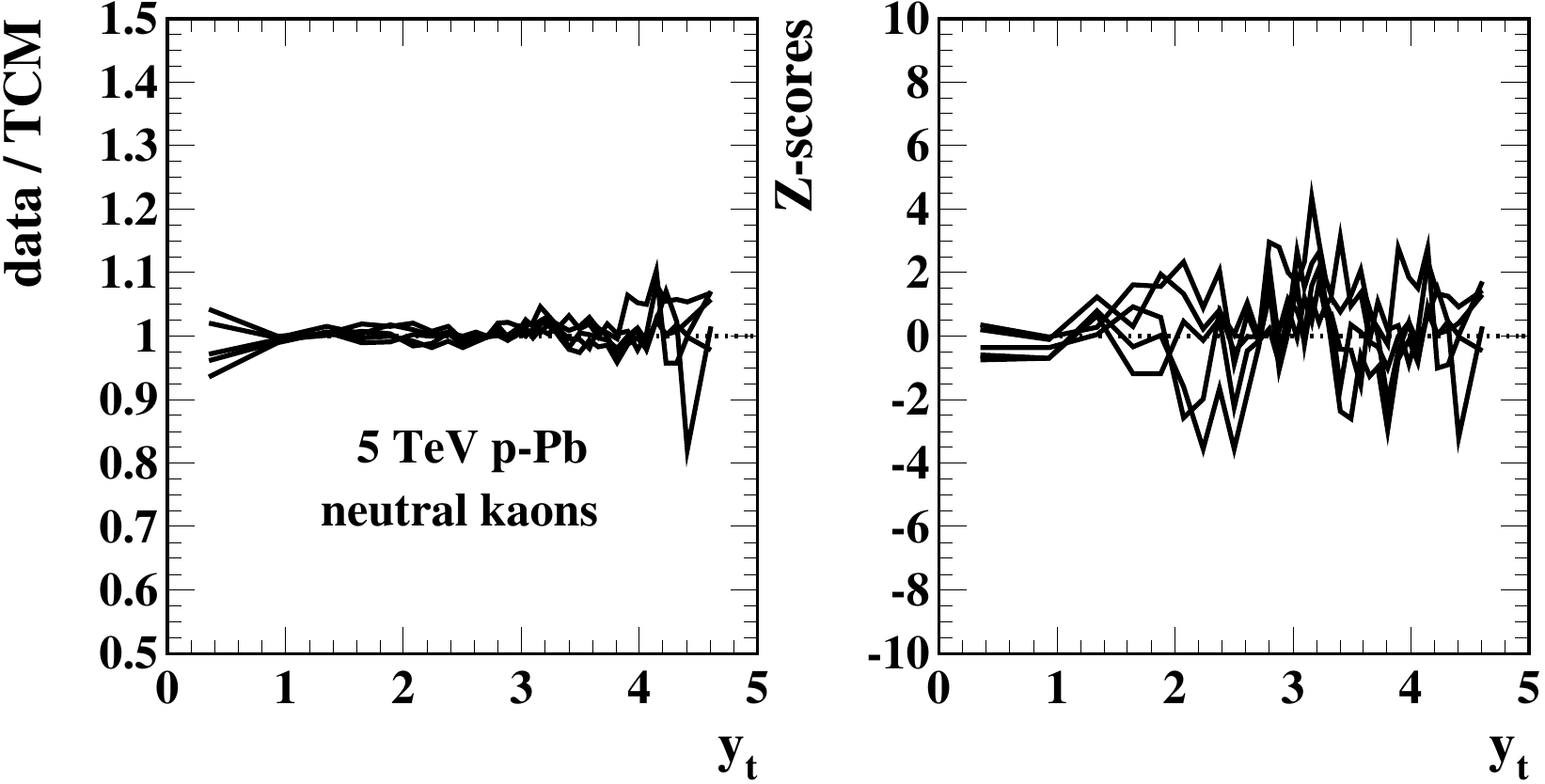}
	\put(-202,105) {\bf (c)}
	\put(-23,105) {\bf (d)}\\
	\caption{\label{zsmeson}
		Left: Data/TCM spectrum ratios for pions (a) and kaons (c).
		Right: 	Z-scores for  pions (b) and kaons (b).
	} 
\end{figure}

\begin{figure}[h]
	\includegraphics[width=3.3in]{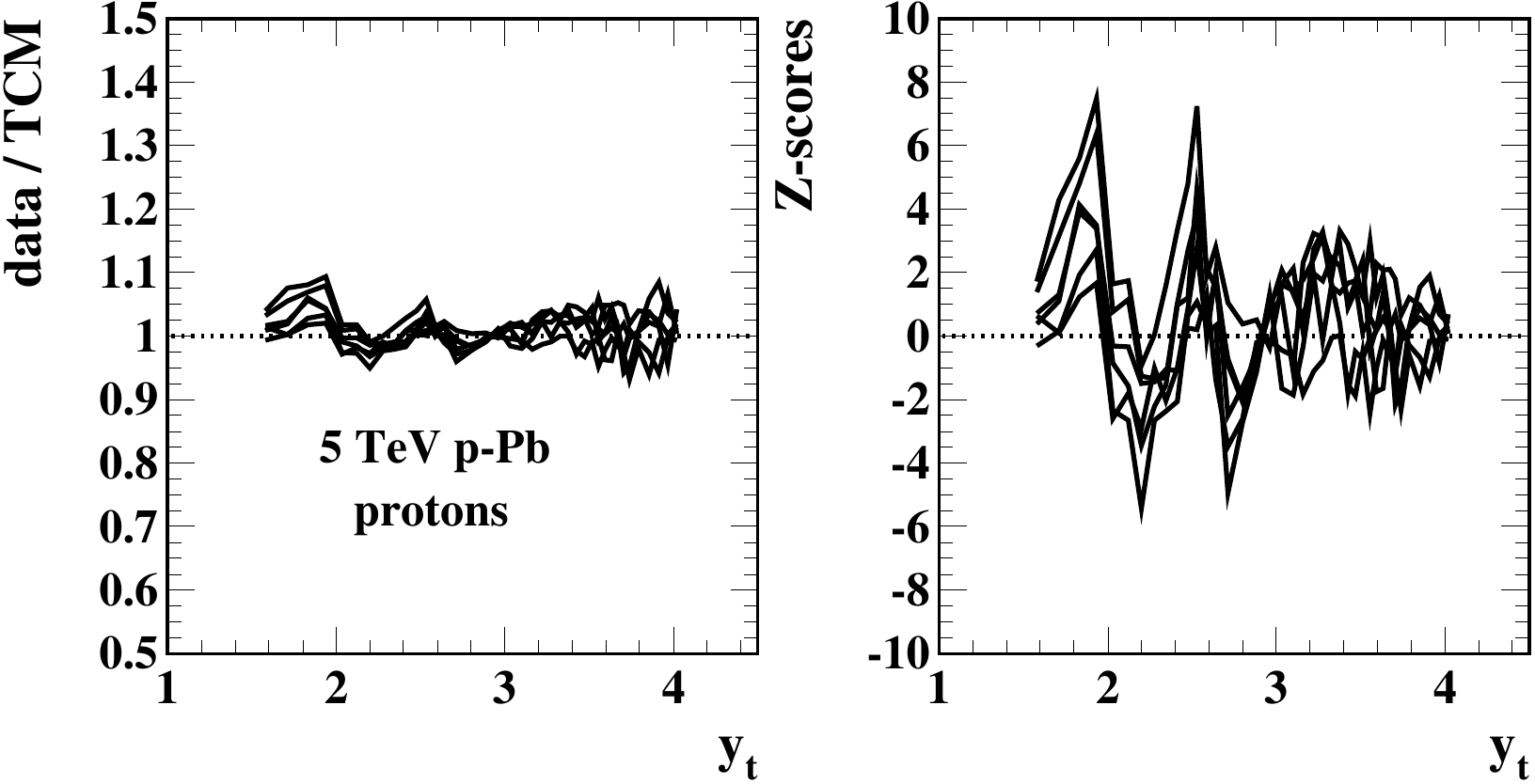}
	\put(-202,105) {\bf (a)}
	\put(-23,105) {\bf (b)}\\
	\includegraphics[width=3.3in]{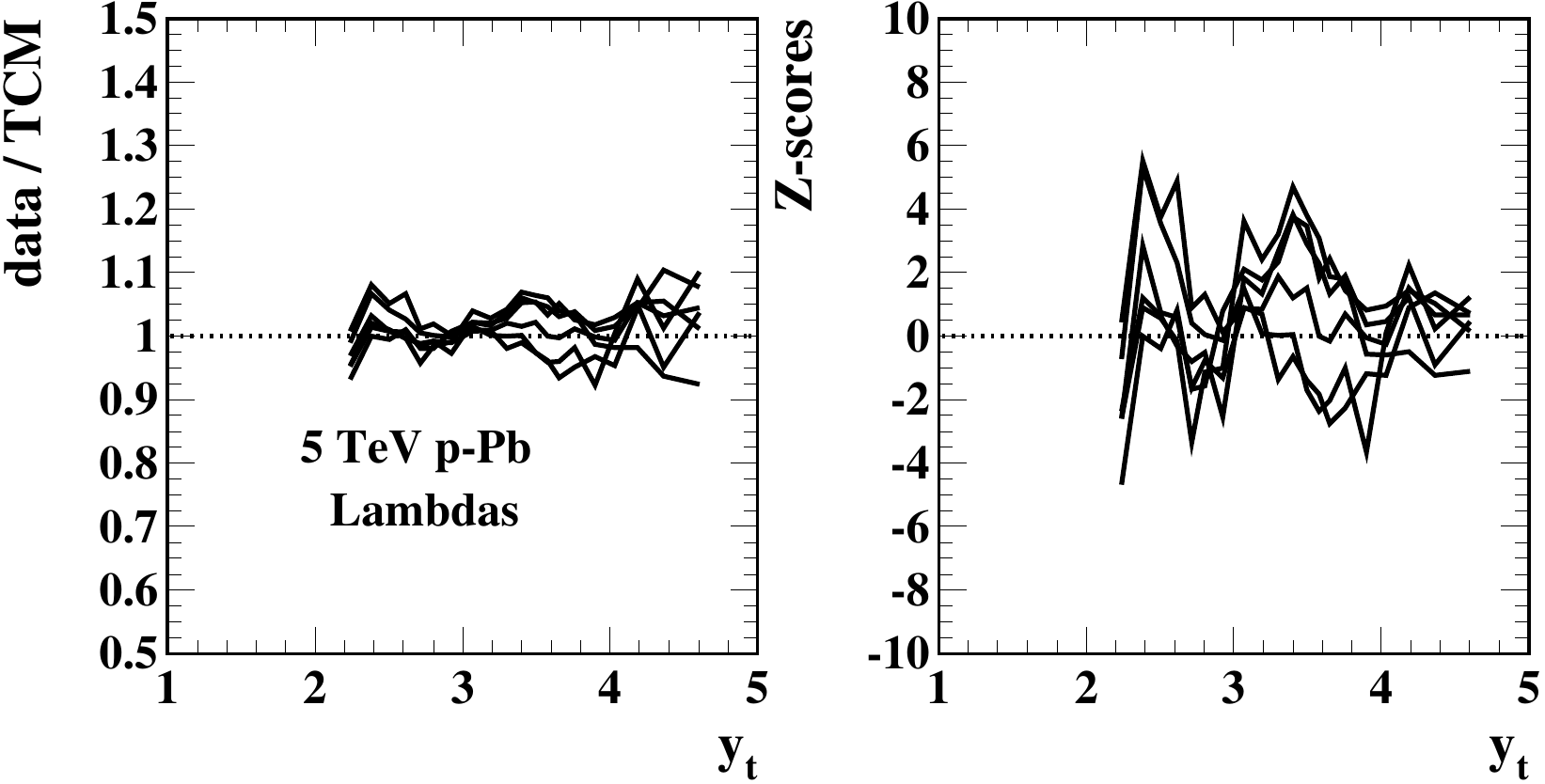}
	\put(-202,105) {\bf (c)}
	\put(-23,105) {\bf (d)}\\
	\caption{\label{zsbaryon}
		Left: Data/TCM spectrum ratios for protons (a) and Lambdas (c).
Right: 	Z-scores for  protons (b) and Lambdas (d).
	}   
\end{figure}

The figures in this subsection compare data/model ratios (left panels) to Z-scores (right panels), for mesons in Fig.~\ref{zsmeson} and for baryons in Fig.~\ref{zsbaryon}. Based on data/model ratios it would seem that the TCM description for pion data is much better than that for Lambda data. Yet a comparison of corresponding Z-scores  shows that the data descriptions are actually of comparable quality.

Examination of Z-scores for four hadron species shows that the r.m.s.\ values are significantly greater than 1. But the excess appears related to substantial systematic deviations (e.g.\ for pions near \yt\ = 2) that cannot be due to some physical phenomenon not accommodated by the TCM. Such deviations more likely arise from some irregularity in processing the primary particle data, as for instance was encountered in Sec.~VI B of Ref.~\cite{tommodeltests}. In general, these Z-scores suggest that the PID TCM represents all physical information carried by \ppb\ \pt\ spectra.

\subsection{TCM defined on the continuum}

The TCM as utilized above is defined on \yt\ data values for direct comparisons with data spectrum components as for figures in this subsection. In following sections the same TCM is defined on a ``continuum'' of 100 equal-spaced values on $y_t \in [0,5]$ ($p_t \in [0,10]$ GeV/c) for investigation of spectrum ratios and ensemble-mean \mmpt\ trends.

The TCM for PID \yt\ spectra configured to accommodate spectrum data as in this section is defined by
\bea \label{pidspectcmxx}
\bar \rho_{0i}(y_t) &=& S_i(y_t) + H_i(y_t) \approx \frac{d^2n_{chi}}{y_t dy_t d\eta}
\\ \nonumber
&=& z_{si}(n_s)  \bar \rho_{s} \hat S_{0i}(y_t) +  z_{hi}(\bar y_t,n_s)  \bar \rho_{h} \hat H_{0i}(y_t,n_s) ,
\eea
where the approximation in the first line arises from averaging charge densities over some limited $\eta$ interval about midrapidity.
The spectrum data reported in Ref.~\cite{aliceppbpid} and appearing in Fig.~\ref{piddata} (as densities on \pt) have the form
\bea
\frac{1}{2\pi} \frac{d^2n_{chi}}{p_t dp_t dy_z} &\approx& \frac{1}{2\pi} \frac{y_t}{m_t p_t} \bar \rho_{0i}(y_t).
\eea
That transformation of $\bar \rho_{0i}(y_t)$ is used to produce the TCM curves in Fig.~\ref{piddata}. Consistent with results above there are no visible deviations between points and curves in Fig.~\ref{piddata} with two exceptions: (a) Meson data hard components for $n = 7$ (most peripheral) fall well below the model Gaussian below the mode leading to a slight displacement of spectrum data below the TCM near \yt\ = 2.7. (b) Baryon data hard components  for $n = 7$ rise above the TCM hard-component model above the mode (apparent in Fig.~\ref{newprot}, a and c) leading to slight elevation of $n = 7$ baryon data above the TCM for large \yt.

\section{$\bf p$-$\bf Pb$ PID spectrum and yield Ratios} \label{specratios}

PID spectrum and yield ratios have often been invoked to argue that hadrons from high-energy nuclear collisions (\pp, \pa\ and \aa) emerge from a partially equilibrated bulk medium via a freeze-out mechanism, thus supporting inference of QGP formation in those systems. Ratio methods from Ref.~\cite{aliceppbpid} are reviewed in the first subsection. The remaining subsections then consider PID spectrum and yield ratios in light of the present TCM analysis.

\subsection{ALICE ratio methods}

Reference~\cite{aliceppbpid} presents PID spectrum and yield ratios in several formats as follows (with figure numbers from that paper).  Figure~2 shows PID spectrum ratios vs \pt\ for three hadron combinations and for two \ppb\ centralities -- 0-5\% and 60-80\% (centralities as reported by Ref.~\cite{aliceppbpid}, see $\sigma' / \sigma_0$ values in Table~\ref{rppbdata}) corresponding to event classes $n = 1$ and 6 in Table~\ref{rppbdata}.  Figure~3 shows integrated-yield $p/\pi$ ratios vs charge density $\bar \rho_0$ for \ppb\ compared to \pbpb\ and inferred power-law exponents $B(p_t)$ (referring to $\bar \rho_0$ dependence) vs integrated \pt\ interval. The latter are obtained for $p/\pi$ and $\Lambda / K_\text{S}^0$. Figure~5 shows integrated-yield ratios for three hadron species to pions vs $\bar \rho_0$ comparing \dau\ and \ppb\ with \auau\ and \pbpb.

Regarding  Fig.~2, baryon/meson ratios ``...show a significant enhancement at intermediate $p_T \sim 3$ GeV/c, qualitatively reminiscent of that measured in Pb-Pb collisions. The latter are generally discussed in terms of collective flow or quark recombination.'' One may ask how ``enhancement'' is measured, relative to what? 

Regarding Fig.~3, ``It is worth noticing that the [yield within \pt\ bins] ratio $p/\pi$ as a function of $dN_{ch}/d\eta$...follows a power-law behavior....the same trend is also observed in \pbpb\ collisions.''
Regarding Fig.~5, ``...a small increase is observed in the $K/\pi$ and $\Lambda/\pi$ ratios.... A similar rise...in \pbpb...collisions...is typically attributed to a reduced canonical suppression of strangeness production...or to an enhanced strangeness production in a quark-gluon plasma.''

Regarding suggestions that certain ratio features may imply QGP formation in small systems one should consider that a simple model describing PID spectrum data, spectrum ratios and yield ratios within their uncertainties appears to require no such {\em ad hoc} mechanisms. The remainder of this section examines ratio trends in the context of a PID TCM for 5 TeV \ppb\ collisions.

\subsection{TCM for PID spectrum ratios}

A TCM for PID spectrum ratios can be derived as follows. The TCM for individual PID spectra in Eq.~(\ref{pidspectcm}) depends critically on accurate values for nonPID \ppb\ geometry parameters as in Table~\ref{rppbdata} which can be used to generate hard/soft ratio $x\nu$. In Sec.~IV of Part I coefficients $z_{si}(n_s)$, $z_{hi}(n_s)$ and ratios $\tilde z_{i}(n_s)$ are determined. In Sec.~\ref{tcmadapt} of this article model functions $\hat H_{0i}(y_t,n_s)$ are optimized to accommodate hard-component variations with \ppb\ centrality. The TCM for a PID spectrum ratio including hadron species $i$ and $j$ is then expressed as
\bea \label{ratiopidtcm}
\frac{\bar \rho_{0i}(y_t;n_s)}{\bar \rho_{0j}(y_t;n_s)} &=&  
\frac{z_{si}(n_s)}{z_{sj}(n_s)} \cdot \frac{\hat S_{0i}(y_t)+ \tilde z_{i}(n_s) x\nu \hat H_{0i}(y_t,n_s)}{\hat S_{0j}(y_t)+\tilde z_{j}(n_s) x\nu\hat H_{0j}(y_t,n_s)}~~
\nonumber \\ 
&\rightarrow& \frac{H_{i}(y_t,n_s)}{H_{j}(y_t,n_s)}
=   \frac{z_{hi}(n_s)}{z_{hj}(n_s)} \cdot \frac{ \hat H_{0i}(y_t,n_s)}{\hat H_{0j}(y_t,n_s)},
\eea
where the second line is the limit for large \yt.

\begin{figure}[h]
	\includegraphics[width=1.67in]{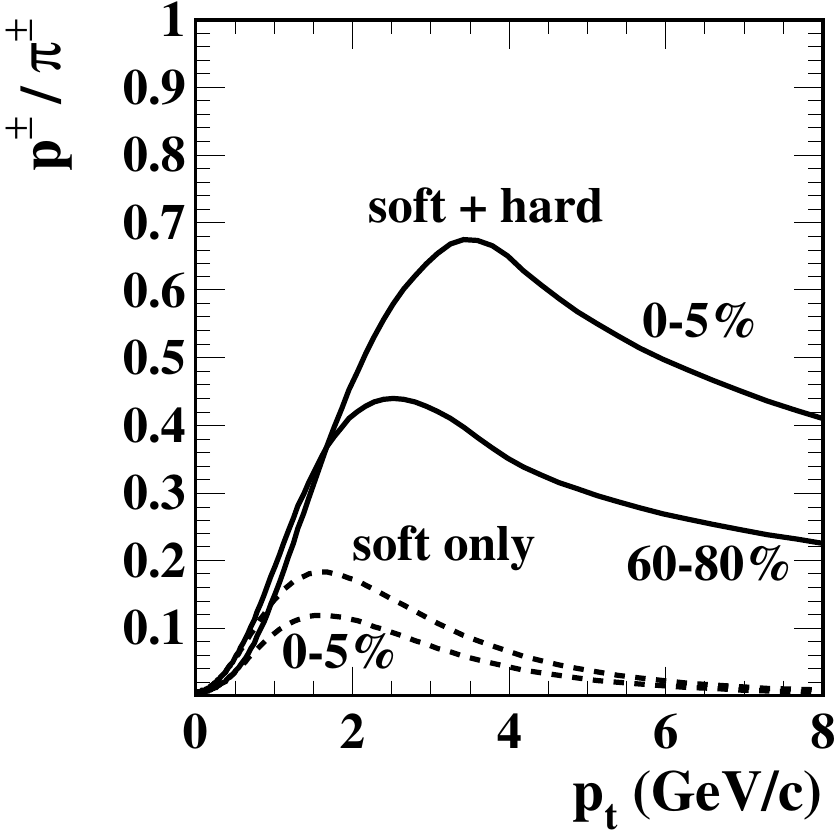}
	\includegraphics[width=1.64in]{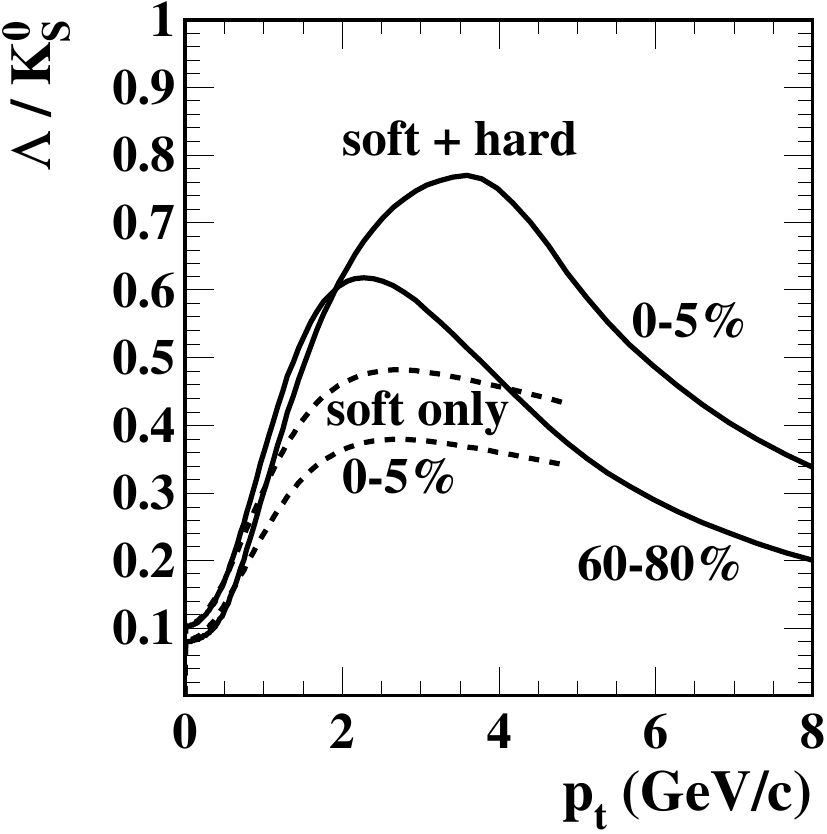}
	\caption{\label{ratios}
		Left: $p/\pi$ spectrum ratios vs \pt\ for two \ppb\ centralities derived from the variable TCM (solid) and from the TCM with soft-component only (dashed). The reversal of centrality dependence near 1.5 GeV/c is due to centroid shifts on \yt\ for protons and pions (in opposite senses) as explained in the text.
		Right:	$\Lambda / K_\text{S}^0$ spectrum ratios treated as in the left panel. $K_\text{S}^0$ spectra are multiplied by 2 corresponding to $\lambda + \bar \lambda$.
}  
\end{figure}

Figure~\ref{ratios} shows spectrum ratios for $p/\pi$ (left) and $\Lambda / K^0_\text{S}$ (right). The general features of the ratio trends are comparable to those in Fig.~2 of Ref.~\cite{aliceppbpid}. However, the data $p/\pi$ trends presented there are 40\% lower than in the left panel of Fig.~\ref{ratios}, consistent with conclusions in Part I.  The $p/\pi$ ratios for peripheral collisions in Ref.~\cite{aliceppbpid} Fig.~2  are quantitatively similar for \ppb\ and \pbpb\ collisions suggesting that the same proton inefficiency noted in Part I occurs for \pbpb\ as well. The ``soft only'' trends in Figure~\ref{ratios} are obtained by omitting the hard components from the TCM and demonstrate in the left panel that above 1 GeV/c ($y_t \approx 2.7$) $p/\pi$ spectrum ratios are dominated by the hard-component (jet) contribution.

An interesting feature of the spectrum ratios is the reversal of sign (asymmetry) of the variation with centrality: Ratios decrease at lower \pt\ with increasing centrality but increase at higher \pt. That feature is emphasized in the top panel of Fig.~3 of Ref.~\cite{aliceppbpid}. The ratio TCM of Eq.~(\ref{ratiopidtcm}) predicts that spectrum ratios should {\em decrease} uniformly on \yt\ with increasing centrality {\em assuming fixed model functions $\hat S_{0i}(y_t)$ and $\hat H_{0i}(y_t)$}. Ratio behavior is then dominated by factors $z_{xi}(n_s)/z_{xj}(n_s)$ ($x = s,\,h$). If species $i$ is more massive $z_{xi}(n_s)$ should decrease more rapidly as noted in Sec.~V of Part I (e.g.\ its  Fig.~9, left).

These results can be compared with Fig.~9 of Ref.~\cite{ppbpid}. The TCM spectrum ratios generated in that case were based on shifting only the baryon hard components, via the {\em ad hoc} shift function described by Eq.~(\ref{adhoc}). However, Sec.~\ref{tcmadapt} of this article establishes that whereas baryon centroids shift to higher \yt\ with increasing centrality meson centroids shift to lower \yt. As discussed below meson centroid shifts should then increase the difference between central and peripheral spectrum ratio trends.

\subsection{Interpreting PID spectrum ratio trends} \label{ratiotrends}

As demonstrated in Sec.~\ref{modelacc} the TCM provides an excellent data description for five hadron species including corrected proton spectra. It is therefore useful to employ the TCM to investigate properties of spectrum ratios and their physical interpretations. There are two main issues: 
(a) the origin of the sign change on \yt\ of the centrality trends and
(b) the origin of the peak near 3 GeV/c.

\begin{figure}[h]
	\includegraphics[width=1.65in,height=1.6in]{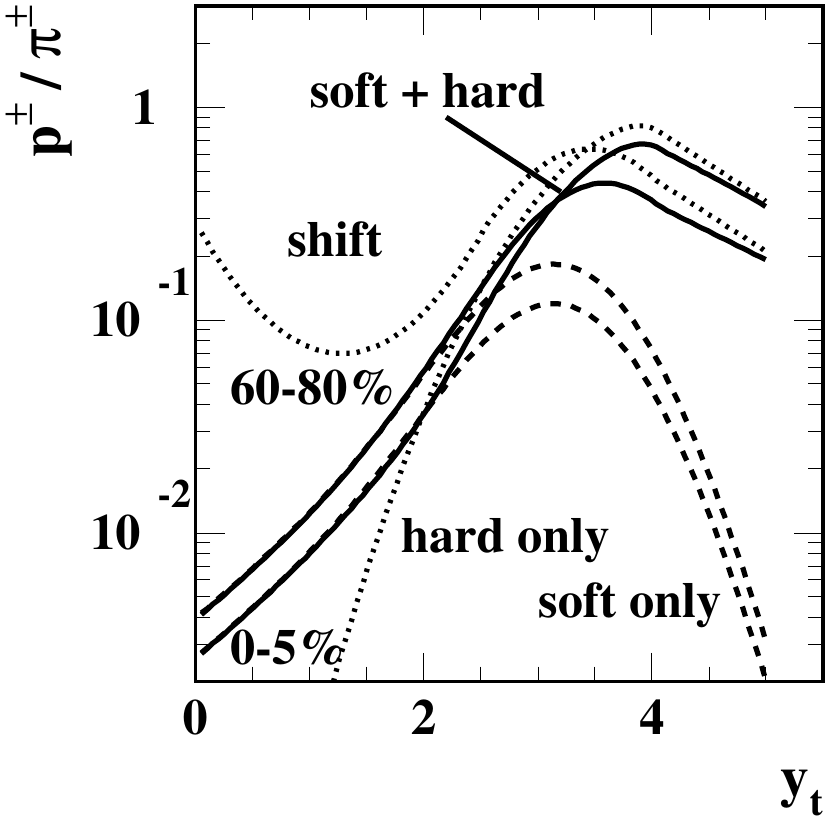}
	\includegraphics[width=1.65in,height=1.6in]{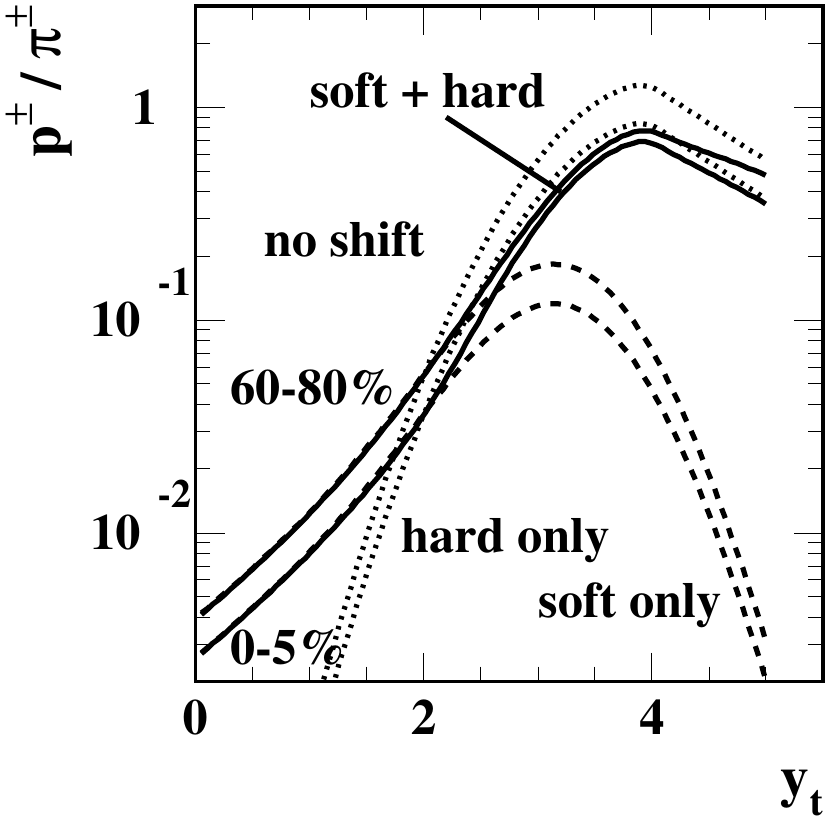}
	\caption{\label{dave}
		Left: Soft+hard spectrum-ratio curves as in Fig.~\ref{ratios} (left) with proton and pion hard components shifted with centrality (solid) and plotted here in a log-log format. The dashed curves are the soft-only curves from Fig.~\ref{ratios} (left). The dotted curves are corresponding hard-only ratios.
		Right:	Same as the left panel except that 
		proton and pion hard components are {\em not} shifted as in Fig.~\ref{ratios}. The reversal of centrality dependence near 1.5 GeV/c ($y_t \approx 3.1$) then does not occur. Note alteration of dotted curves with and without shifting.
}  
\end{figure}

Figure~\ref{dave} shows $p/\pi$ spectrum ratios for TCM PID spectra, transforming  the usual linear spectrum ratio vs \pt\ to what is effectively a log-log representation. The log-log format permits more comprehensive examination of data trends and isolation of jet and nonjet contributions. The solid and dashed curves in the left panel  are the same as those in Fig.~\ref{ratios} (left). Reversal in sign of centrality evolution below and above $y_t \approx 3.1$ in Fig.~\ref{dave} (left) reflects PID spectrum data accurately modeled via the PID TCM to their statistical limits as described in Sec.~\ref{tcmadapt}. As noted above, the sign change results from shifts of hard-component centroids to higher \yt\ for baryons (numerators) and to lower \yt\ for mesons (denominators).

The effect of hard-component centroid shifts for two hadron species is illustrated by the dotted curves in Fig.~\ref{dave} (left). Note that the most dramatic \nch\ variation in the $p/\pi$ hard-component ratio occurs for $y_t < 2$ ($p_t < 0.5$ GeV/c). Above that point the ratio hard component for 60-80\% central is shifted to lower \yt\ and lower amplitude via proton and pion centroid shifts compared to the 0-5\% central curve corresponding to $n =1$ reference model functions. The combination results in the trend reversal.

Figure~\ref{dave} (right) shows TCM spectrum ratios with no centroid shifts. Note that the 0-5\% curves are the same in the two panels by construction. The ratios (and individual soft and hard components) then {\em increase} with {\em decreasing} centrality consistently on \yt\ per Eq.~(\ref{ratiopidtcm}). In Ref.~\cite{ppbpid} the proton hard-component model was shifted to illustrate this point in its Fig.~9. The pion shift to {\em lower} \yt\ as in Fig.~\ref{newpion} (left) above, acting in the $p/\pi$ {\em denominator}, was expected to significantly increase the asymmetry. Results for proton and pion peak shifts separately as presented in the following figure confirm that expectation.

Figure~\ref{davex} shows $p/\pi$ spectrum ratios (solid) where only the pion hard component shifts with centrality (left) or only the proton hard component shifts (right). The bold dotted curves in this figure are solid curves in Fig.~\ref{ratios} (left) representing observed data trends. The curves for 0-5\% are unchanged by construction (corresponding to reference hard-component model functions with no centroid shifts). Compared to the 60-80\% curve (upper dotted) from Fig.~\ref{dave} (right) the corresponding 60-80\% curves in Fig.~\ref{davex} (solid) have translated about half way toward the 60-80\% curve (lower dotted)  from Fig.~\ref{ratios} (left) representing data. These result demonstrate that proton centroid shifts to {\em higher} \yt\ and pion shifts to {\em lower} \yt\ contribute similarly to sign reversal for the $p/\pi$ spectrum ratio.

\begin{figure}[h]
	\includegraphics[width=1.65in,height=1.6in]{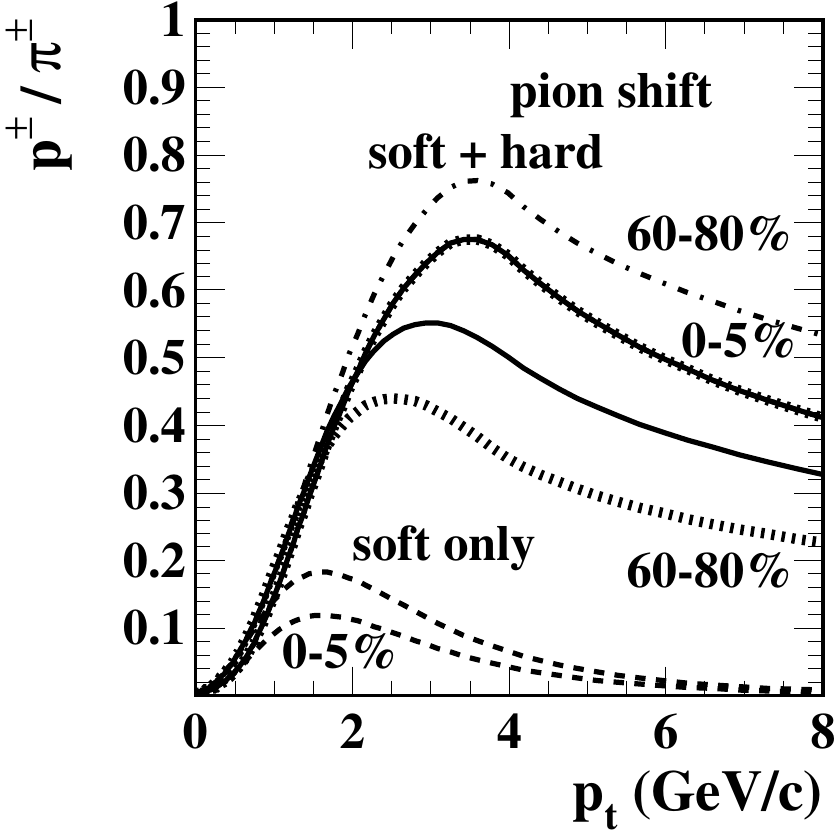}
	\includegraphics[width=1.65in,height=1.6in]{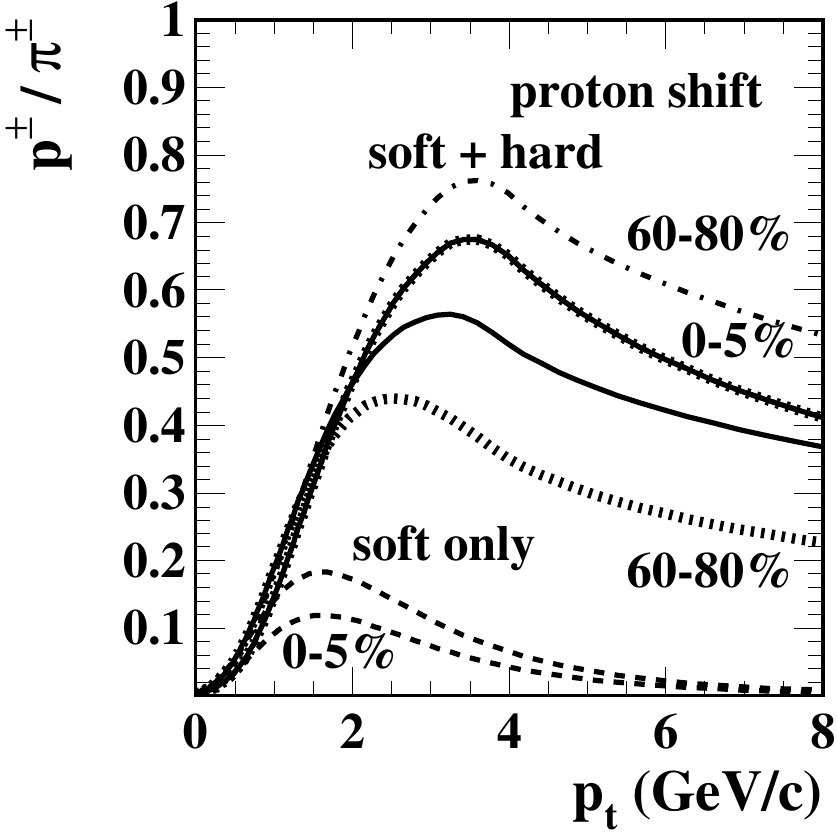}
	\caption{\label{davex}
		$p/\pi$ spectrum TCM ratios with pion centroid shift only (left) and proton centroid shift only (right). The dash-dotted curves correspond to a 60-80\% model with no centroid shifts as reference. The bold dotted curves are the solid curves in Fig.~\ref{ratios} (left) that describe data accurately. The lower solid curves are 60-80\% model curves with pion {\em or} proton centroid shifts. The upper solid curves indicate that the 0-5\% model (reference) is independent of centroid shifts by construction.
	} 
\end{figure}

The general structure of the spectrum ratios, and especially the prominent peak near 3 GeV/c, are explained as follows: The soft-only trends (dashed curves above) arise from interplay of slope parameter $T$ and L\'evy exponent $n$ in $\hat S_0(y_t)$. At $y_t =1$ ($p_t \approx 0.15$ GeV/c) the soft $p/\pi$ ratio is small $O(0.01)$ reflecting statistical-model hadronic abundances. However, the soft ratio increases strongly because the proton slope parameter 210 MeV is substantially greater than the pion value 145 MeV. Above 1 GeV/c ($y_t \approx 2.7$) the lower pion L\'evy exponent $n= 8.5$ (harder tail) then overwhelms the proton value 14 and the $p/\pi$ soft trend drops rapidly. Above 1 GeV/c ($y_t \approx 2.7$) pion and proton hard components dominate.

The hard-only trends (dotted) in Fig.~\ref{dave} are similarly explained. The pion hard-component centroid $y_t \approx 2.46$ is substantially lower than the proton centroid  $y_t \approx 2.99$ and the pion peak is broader. The pion hard component is thus much greater than the proton hard component below their modes. But as \yt\ traverses the peak modes the proton hard component comes to dominate near and above its 3 GeV/c mode, producing a peak there in the $p/\pi$ ratio. Above 4 GeV/c ($y_t \approx 4$) the difference in peak widths and exponential (power-law) tail constants $q$ prevails. The pion $q \approx 4$ (harder tail) vs the proton $q \approx 5$ (softer tail) insures that the $p/\pi$ hard-component ratio then falls off strongly. As is evident from Fig.~\ref{ratios} (left) the structure of the $p/\pi$ ratio on \yt\ is dominated by jet-related spectrum hard components above 1 GeV/c.

\subsection{Comparing $\bf p$-Pb ratios to A-A spectrum ratios}

In Fig.~2 of Ref.~\cite{aliceppbpid} PID spectrum ratios for 5 TeV \ppb\ collisions are compared with those from 2.76 TeV \pbpb\ collisions. The ratio structure, including the asymmetry on \yt\ with increasing centrality, is formally similar although the effect is much larger for the \pbpb\ data. The ratio peak near 3 GeV/c in \pbpb\ is said to be ``generally discussed in terms of collective flow or quark recombination.'' It is noted that for a \pbpb\ centrality bin with matching ratio peak amplitude the corresponding \pbpb\ $\bar \rho_0$ is 1.7 {\em higher} than that for central \ppb.

\begin{figure}[h]
	\includegraphics[width=1.65in,height=1.6in]{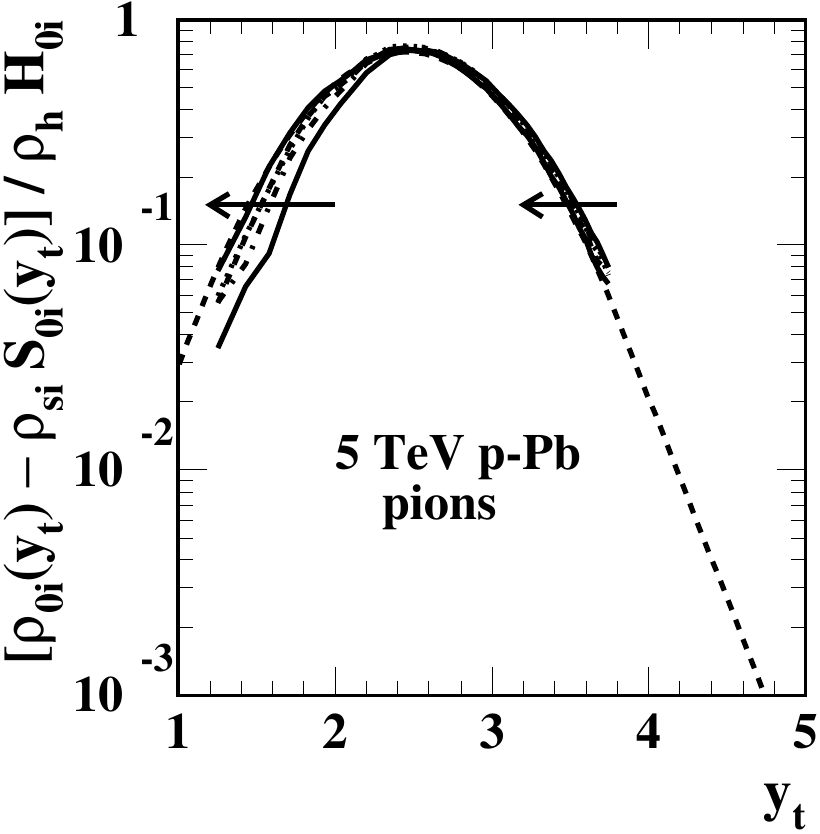}
	\includegraphics[width=1.65in]{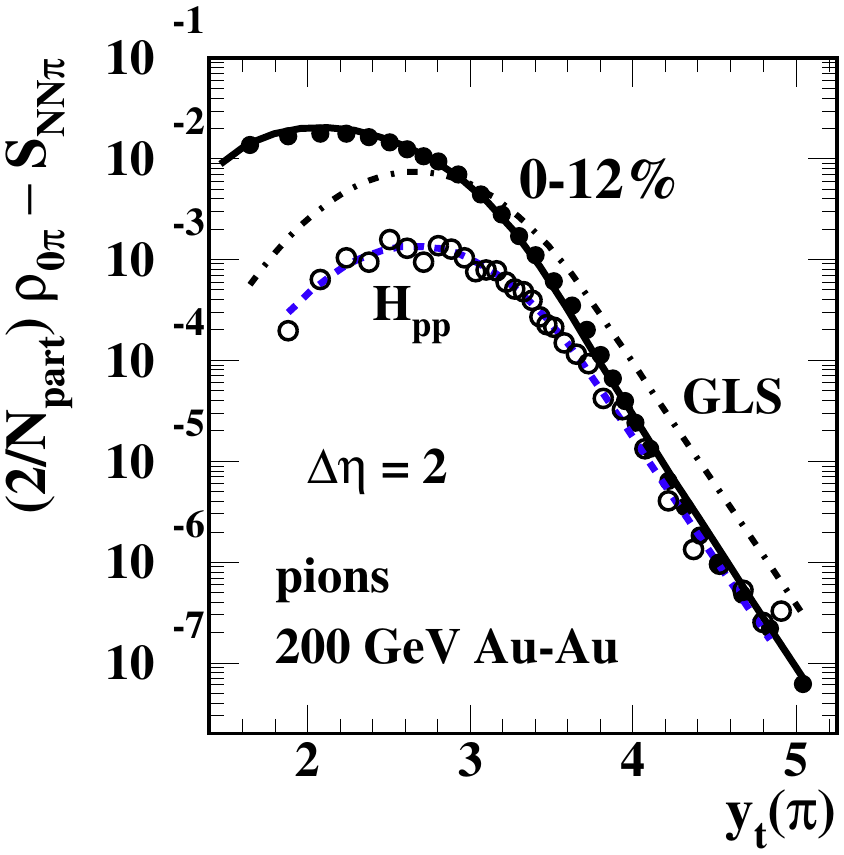}
	\caption{\label{comparerats}
		Left: Repeat of Fig.~11 (left) of Part I showing shift of pion hard-component centroid to {\em lower} \yt\ with increasing \ppb\ centrality.
		Right:	Pion hard components for 200 GeV \pp\ (open circles) and 0-12\% central \auau\ (solid dots) (Fig.~18, right of Ref.~\cite{pbpbpid}). The TCM reference for central \auau\ is the dash-dotted curve -- the TCM for minimum-bias \pp\ (dashed) multiplied by $\nu \approx 6$. Comparison of \auau\ central pion data with the TCM reference reveals a centroid shift to lower \yt\ much larger than that for \ppb\ at left.
} 
\end{figure}

Figure~\ref{comparerats} (left) repeats Fig.~11 (left) of Part I showing the pion hard-component centroid shifted to lower \yt\ with increasing \ppb\ centrality. It is demonstrated in the previous subsection that such displacements are responsible for the asymmetry on \yt\ of PID spectrum ratios with increasing centrality, where the same asymmetry effect arises if the ratio numerator is shifted to higher \yt\ or the denominator is shifted to lower \yt. How then should that relate to the \pbpb\ results appearing in Ref.~\cite{aliceppbpid}?

Figure~\ref{comparerats} (right) shows PID pion spectrum hard components for 200 GeV minimum-bias (MB) \pp\ collisions (open circles) and 0-12\% central 200 GeV \auau\ collisions (solid dots). The dashed curve is the TCM for 200 GeV MB \pp\ collisions while the dash-dotted curve shows the TCM {\em prediction} for central \auau\ collisions assuming Glauber linear superposition (GLS) -- the dashed curve scaled up by $\nu =$ 6 as the ratio of N-N binary collisions to participant pairs for central \auau. The important result: the pion {\em data} hard component is shifted to lower \yt\ with increasing \pbpb\ centrality compared to linear superposition, formally equivalent to the shift in the left panel. It is also significant that the effective peak shift is correlated with {\em suppression} at higher \yt\ that is usually interpreted to demonstrate ``jet quenching.''

Given this \auau\ example it is reasonable to interpret the \pbpb\ data in Fig.~2 (middle) of Ref.~\cite{aliceppbpid} as arising from some combination of proton hard component shifting to higher \yt\ and pion hard component shifting to lower \yt\ (as in Fig.~\ref{comparerats}, right). The correspondence with a spectrum feature associated with jet quenching buttresses the conclusion that the most prominent features of spectrum ratios and their centrality evolution are mainly jet related. It should also be noted that $p/\pi$ spectrum ratios for peripheral collisions in Fig.~2 of Ref.~\cite{aliceppbpid} are the same within uncertainties for \ppb\ and \pbpb\ collisions, suggesting that the proton detection inefficiency evaluated for \ppb\ collisions in Part I is also present in the \pbpb\ data presented in Fig.~2 of Ref.~\cite{aliceppbpid}. See Ref.~\cite{alicepbpbpidspec} for  PID spectrum TCM analysis methods applied to 2.76 TeV \pbpb\ spectrum data. Reference~\cite{aliceppbpid} also states that ``The separation power achieved in p-Pb collisions is identical to that for pp collisions'' suggesting a common property of  any ALICE PID spectrum data.



\subsection{Power-law behavior of PID spectrum ratios}

In its Fig.~3 Ref.~\cite{aliceppbpid} represents the asymmetry on \yt\ of spectrum ratios as in Fig.~\ref{ratios} in terms of power-law exponent $B(p_t;\bar \rho_0)$, i.e.\ $p/\pi \propto \bar \rho_0^{\,B(p_t;\bar \rho_0)}$. In Ref.~\cite{aliceppbpid} values of $B(p_t;\bar \rho_0)$ for successive values of \pt\ are apparently derived via power-law fits to ratio trends on \nch. The same result can also be achieved via logarithmic derivatives:
\bea \label{powerr}
B(y_t;\bar \rho_0) &\approx& \frac{d}{d\ln \bar \rho_0} \ln\left[ \frac{\bar \rho_{0p}(y_t;\bar \rho_0)}{\bar \rho_{0\pi}(y_t;\bar \rho_0)} \right]
\\ \nonumber
&\approx& B_p(y_t;\bar \rho_0) - B_\pi(y_t;\bar \rho_0),
\eea
where the switch to transverse rapidity \yt\ facilitates the discussion below. The logarithm of a ratio results in a difference between two hadron species represented by the second line. Results in Fig.~\ref{ratios} (left) make clear that near the centrality-trend reversal (asymmetry) of the $p/\pi $ ratio hard components $H_{i}(y_t)$ dominate. Also, in the neighborhood of the $p/\pi$ ratio peak the hard-component peaks are approximately Gaussian with mean $\bar y_t$ and width $\sigma_{y_t}$ leading to the following progression for species $i$:
\bea \label{powerb}
B_i(y_t;\bar \rho_0) &\approx&  \bar \rho_0\frac{d}{d \bar \rho_0} \ln\left[ {H_{i}(y_t;\bar y_t, \bar \rho_0)} \right]
\\ \nonumber
&=&  \bar  \rho_0\frac{d\bar y_t}{d \bar \rho_0} \cdot \frac{d}{d\bar y_t} \ln\left[ {H_{i}(y_t;\bar y_t, \bar \rho_0)} \right]
\\ \nonumber
&\approx&  \bar  \rho_0\frac{d\bar y_t}{dn} \frac{dn}{d \bar \rho_0} \cdot (y_t - \bar y_{tn}) / \sigma^2_{y_t}.
\eea
For five of seven \ppb\ centrality classes $n \in [2,6]$ the expression in the first line is evaluated in terms of finite differences between data hard components for four nearest-neighbor centrality ($\bar \rho_0$) pairs, and the results are averaged. Given the last line above, based on a Gaussian model, the averaged data trend is approximated as
\bea \label{bi}
\bar B_i(y_t) &\approx&   \hat B_i  (y_t - \bar y_t).
\eea

\begin{figure}[h]
	\includegraphics[width=1.65in,height=1.6in]{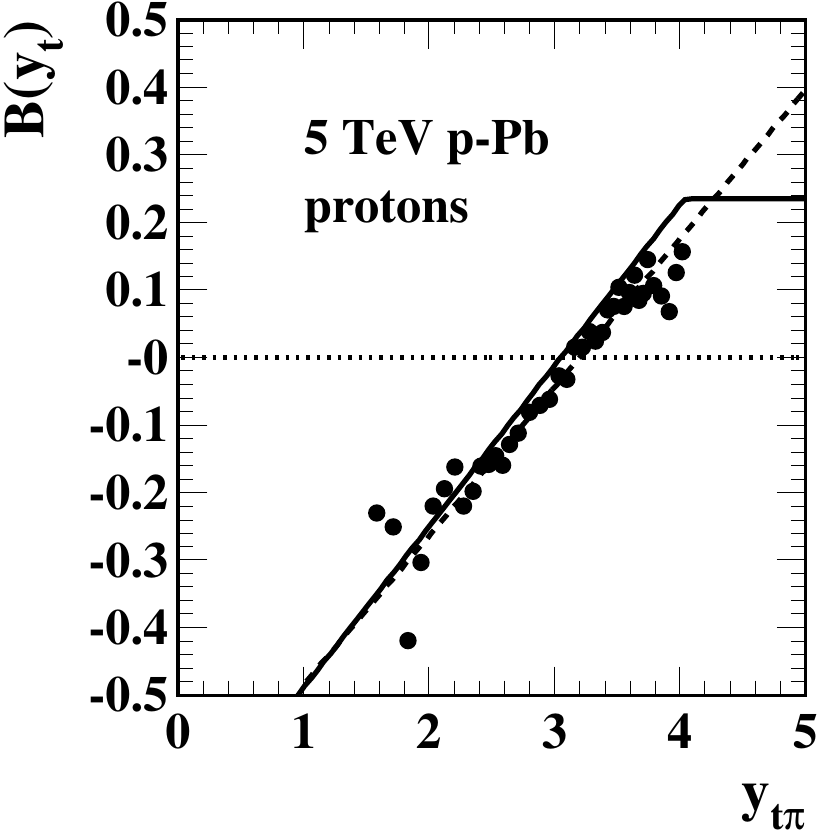}
	\includegraphics[width=1.65in,height=1.6in]{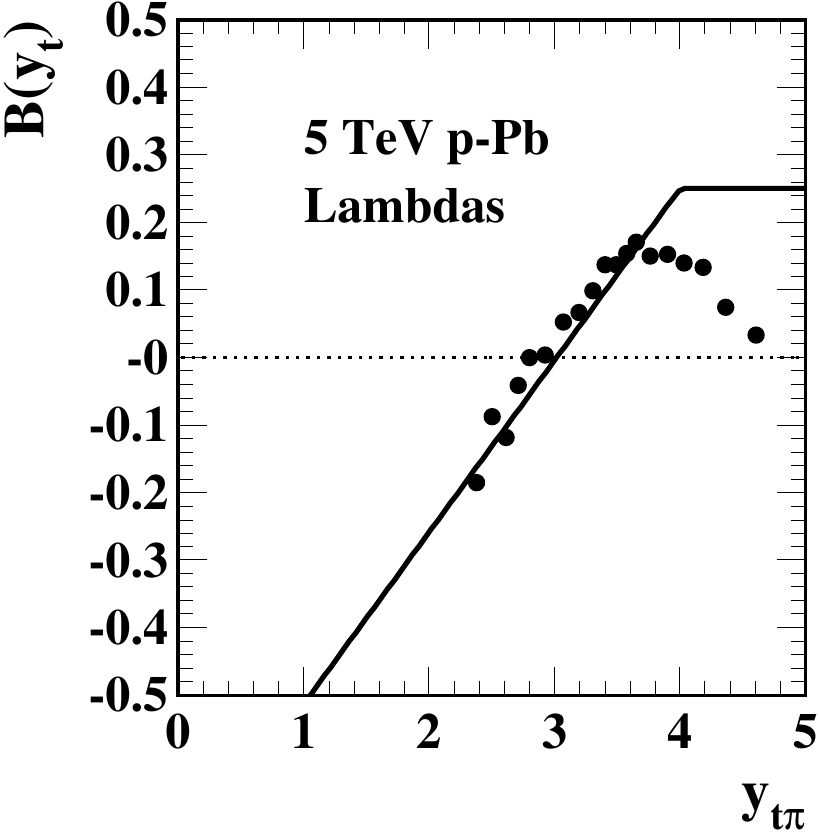}
	\caption{\label{xxx}
	Power-law exponent $B(y_t)$ (points) obtained from data hard components in Fig.~\ref{newprot} (a) via Eq.~(\ref{powerb}) (first line). Solid curves are derived from the same procedure applied to TCM hard components $\hat H_0(y_t,n_s)$ shifted on \yt\ with $n_s$ as described in Ref.~\cite{ppbpid} Sec.~6.2. Note that $y_t \approx 3.1$ above is equivalent to $p_t \approx 1.55$ GeV/c in Fig.~3 (middle) of Ref.~\cite{aliceppbpid}.
}  
\end{figure}

Figure~\ref{xxx} shows inferred data power-law exponents (points) for protons (left) and Lambdas (right). The solid curves correspond to Eq.~(\ref{powerb}) applied to TCM model function $\hat H_0(y_t;n_{s})$ shifted by $\Delta \bar y_t = 0.015$ for each increment in centrality index $n$ (Table~\ref{rppbdata}) as per Ref.~\cite{ppbpid}. The horizontal line segment corresponds to the exponential (power-law) tail on $\hat H_0(y_t,n_{s})$. Note that  the Lambda data trend extends high enough on \yt\ to reveal that the data exponential {\em tail} for baryons does not shift significantly with centrality, as is evident from Fig.~\ref{newprot}.

The slope of the $\bar B_i(y_t)$ trend can be predicted as follows (see Table~\ref{rppbdata}): Assume centrality class $n = 4$ with $d\bar y_t / dn \approx 0.015$ as noted above, $\bar \rho_0 \approx 23$, $d\bar \rho_0 / dn \approx 7$ and $\sigma_{y_t} \approx 0.47$. Based on Eq.~(\ref{powerb}) (third line) baryon $\hat B_i \approx 0.22$ in Eq.~(\ref{bi}) corresponding to the trends in Fig.~\ref{xxx}. The dashed curve at left is $\bar B_p(y_t) = 0.22(y_t - 3.2)$. 

For the full $p/\pi$ ratio the pion contribution $B_\pi(y_t)$ in Eq.~(\ref{powerr}) must contribute. The denominator should nominally contribute a negative term as in Eq.~(\ref{powerr}). However, as shown in Fig.~\ref{newpion} the factor $d\bar y_t/dn$ for pions is itself negative (the hard component shifts to {\em lower} \yt) so the result should be further increase in the slope of total $B(y_t)$. Those relations are visualized in Fig.~\ref{davex}.

As demonstrated above, various features of Figs.~2 and 3 in Ref.~\cite{aliceppbpid} are actually manifestations of jet-related PID spectrum hard components shifting to lower or higher \yt\ with increasing \nch. The chosen plotting formats in Ref.~\cite{aliceppbpid} in effect obscure the consequences of jet production in \ppb\ collisions as reported in Sec.~\ref{tcmadapt} above. 

\subsection{PID yield ratios}

In its Fig.~5 Ref.~\cite{aliceppbpid} presents centrality evolution for three hadron yield ratios from 5 TeV \ppb\ and various \aa\ collision systems. According to Eq.~(\ref{ratiopidtcm}) (first line) integration of numerator and denominator leads to
\bea \label{pidyieldrat}
\frac{\bar \rho_{0i}(n_s)}{\bar \rho_{0j}(n_s)} &=& \frac{z_{si}(n_s)}{z_{sj}(n_s)} \cdot \frac{1+ \tilde z_{i}(n_s) x\nu }{1+\tilde z_{j}(n_s) x\nu}
\\ \nonumber
&\rightarrow& \frac{z_{0i}}{z_{0j}}
\eea
as the TCM for yield ratios. The second line of Eq.~(\ref{pidyieldrat}) is a trivial result given the definition of $z_{si}(n_s)$. The TCM that describes PID \ppb\ spectra within statistical uncertainties (see Sec.~\ref{modelacc}) thus {\em predicts} yield ratios independent of \nch. The approximate centrality independence of parameters $z_{0i}$ is confirmed in Sec.~V B of Part I.

The constant values predicted by the TCM (see Table~\ref{pidparamsxx} for $z_{0i}$ estimates) are $K/\pi \approx 0.128 / 0.82 = 0.156$, $p/\pi \approx 0.065 / 0.82 = 0.079$ and $\Lambda / \pi \approx 0.034 / 0.82 = 0.041$. Those values can be compared with the ratio trends in Fig.~5 of Ref.~\cite{aliceppbpid}. For $K/\pi$ the reported ratios increase from 0.12 to 0.145. For $p/\pi$ the reported ratios are approximately constant at 0.057. For $\Lambda / \pi$ the reported ratios increase from 0.032 to 0.040. Given the large {\em overall} systematic uncertainties the $K/\pi$ and $\Lambda / \pi$ increases might not be significant. However, the uncertainties ``uncorrelated across multiplicity bins'' (shaded boxes) are small enough to suggest that those variations are quite significant. Certainly significant is the discrepancy between the $p/\pi \approx 0.079$ TCM prediction and the reported 0.057 ratio in Fig.~5 of Ref.~\cite{aliceppbpid} that would suggest a proton $z_{0i} \approx 0.057 \cdot 0.82 = 0.047 \rightarrow 0.72 \times 0.065$.

The relation between $p/\pi$ and $\Lambda / \pi$ ratios is illuminated by a comparison of proton and Lambda spectra in Fig.~7 of Part I. 
{\em Uncorrected} (i.e.\ published) proton data below 0.6 GeV/c ($y_t \approx 2.15$) divided by 2 agree with corresponding Lambda data within a few percent. That result strongly argues that the $p/\pi$ ratio should be approximately twice the $\Lambda / \pi$ ratio $\approx 0.04$. The reported ratio data $p/\pi \approx 0.057$ ($\approx 70$\% of the expected 0.079) buttresses the correction procedure reported in Part I.

Additional information comes from consideration of sum rules applied to {\em charged} hadrons, particularly the condition  $\sum_i z_{si}(n_s) \approx 1$, with the limiting case (for $n_s \rightarrow 0$) $\sum_i z_{0i} \approx 1$. Note that the $z_{si}(n_s)$ are inferred from published data spectra at low \yt\ where no data require correction and this sum rule is accurately satisfied. The sum rule can then be used to describe a relation among ratios of charged hadrons: 
\bea
K/\pi + p/\pi &\approx& 1/\pi - 1
\eea
where the respective $z_{0i}$  are represented by particle symbols. Assuming that $z_{0i}$ for pions is $\approx 0.8$ within a few percent then one should expect $K/\pi + p/\pi \approx 0.25$, and the sum for TCM ratios is $\approx$ 0.24. But the sum of corresponding ratio data in Fig.~5 of Ref.~~\cite{aliceppbpid} varies with centrality from 0.18 to 0.20. The shortfall of $\approx 0.04$ - 0.06 is highly significant compared to the uncertainties ``uncorrelated across multiplicity bins'' of $\approx 0.002$ reported in Ref.~~\cite{aliceppbpid}. Part of that shortfall (0.022) appears to be a $\approx 30$\% reduction in the {\em total} integrated proton yield, or a larger fraction above 0.6 GeV/c as noted in Part I. Another substantial part (0.021) is an apparent $\approx 15$\% reduction in the K$^{\pm}$ yield (averaged over centrality), possibly due to extrapolation from a limited \pt\ acceptance (see large error bars on charged-kaon \mmpt\ data in Fig.~\ref{rrr}).

\section{Ensemble-mean $\bf \bar p_t$ trends} \label{pidmmpt}

The updated TCM for PID spectra presented in the previous sections may be tested by comparison with measured PID \mmpt\ values. In this section the \mmpt\ TCM is generalized to include $\bar p_{ti}$ hard-components $\bar p_{thi}(n_s)$ and ratios $\tilde z_i(n_s) = z_{hi}(n_s) / z_{si}(n_s)$ varying with centrality assuming that all hadron species share common \ppb\ centrality parameters $x(n_s)$ and $\nu(n_s)$ presented in Table~\ref{rppbdata}. 

\subsection{Ensemble-mean $\bf \bar p_t$ TCM}

The ensemble-mean {\em total} \pt\ $\bar P_{ti}$ for identified hadrons of species $i$ integrated over some angular acceptance $\Delta \eta$ as derived from Eq.~(\ref{pidspectcm}) (second line) is
\bea \label{ptintid}
\bar P_{ti} &=& \Delta \eta \int_0^\infty dp_t\, p_t^2\, \bar \rho_{0i}(p_t)
= \bar P_{tsi} + \bar P_{thi}~~
\\ \nonumber
&=& z_{si}(n_{s})  n_{s} \bar p_{tsi} + z_{hi}(n_s) n_{h} \bar p_{thi}(n_s).
\eea 
Based on spectrum data, and just as for unidentified hadrons, it is concluded that soft-component $\bar p_{tsi}$ is a universal quantity for each hadron species $i$. Hard-component $\bar p_{thi}(n_s)$ trends are inferred from spectrum hard components as they are described in Sec.~\ref{tcmadapt}. 
The corresponding TCM for conventional ratio $\bar p_{ti}$  is
\bea \label{pampttcmid}
\frac{\bar P_{ti}}{\bar n_{chi}} &\equiv&   \bar p_{ti}(n_s)   \approx \frac{\bar p_{tsi} + \tilde z_i(n_s) x(n_s) \nu(n_s) \, \bar p_{thi}(n_s)}{1 +  \tilde z_i(n_s)x(n_s)\, \nu(n_s)},~~~~
\eea
assuming that \pt\ integrals extend down to $p_t = 0$ (i.e.\ by extrapolation of $\bar \rho_{0i}$ data). The lower limit on centrality for $\bar p_{ti}$ is then $\bar p_{tsi}$. The limit for very large multiplicity is the limiting value for $\bar p_{thi}(n_s)$. Data trends and corresponding TCM are illustrated in Fig.~8 of Ref.~\cite{ppbpid}.

In Ref.~\cite{ppbpid} ratio $\tilde z_i = z_{hi} / z_{si}$ for each hadron species was held fixed independent of centrality with values given by its Table 4. That table also includes fixed values for $\bar p_{tsi}$ and $\bar p_{th0i}$ derived from fixed model functions $\hat S_{0i}(y_t)$ and $\hat H_{0i}(y_t)$  respectively as defined by parameters in its Table~2. Centrality parameters $x(n_s)$ and $\nu(n_s)$, independent of hadron species, correspond to Table~\ref{rppbdata} of the present article. In this study ratios  $\tilde z_i(n_s)$ are taken from Sec.~V of Part I, and as noted $\bar p_{thi}(n_s)$ are derived from varying data hard components as derived in Sec.~\ref{tcmadapt}.

Soft-components $\bar p_{tsi}$ are obtained from model functions $\hat S_0(y_t)$ for pions, kaons, protons and Lambdas as $0.40 \pm 0.02$, $0.60 \pm 0.02$, $0.74 \pm 0.02$ and $0.78 \pm 0.02$ GeV/c.

\subsection{Hard components $\bf \bar p_{thi}(n_s)$}

Table~\ref{hardpt} presents hard-component $\bar p_{thi}(n_s)$ values for four hadron species obtained by evaluating sums over hard-component data {\em on data points} (no extrapolation). Proton and Lambda TCM model functions are identical. What sets the proton {\em data} apart is an apparent excess on the low-\yt\ side of the hard-component peak as seen in Fig.~13 (a) of Part I. Results for $K^\pm$ are unreliable due to the limited \pt\ coverage. $K_\text{S}^0$ values are used instead.

\begin{table}[h]
	\caption{Spectrum hard-component means $\bar p_{thi}(n_s)$ for four hadron species from six centralities of 5 GeV \ppb\ collisions with units GeV/c. $K^\pm$ values are copied from $K_\text{S}^0$ values because of the latter's much larger \yt\ coverage. These value are obtained by integrating over data hard components within their \pt\ acceptances.
	}
	\label{hardpt}
	\begin{center}
		\begin{tabular}{|c|c|c|c|c|c|} \hline
			$n$	& $ \pi^\pm $  & 	$K^\pm$  & 	$p$  & 	$K_\text{S}^0$ & 	$\Lambda$  \\ \hline
			1  & $ 1.05$  & -- & $ 1.60 $ & $1.34 $ & $ 1.72 $  \\ \hline
			2  & $ 1.07$  &  -- & $ 1.58 $ & $ 1.35$ & $ 1.71 $  \\ \hline
			3  & $ 1.07 $  &  --  & $ 1.59 $ & $ 1.38$ & $ 1.70 $  \\ \hline
			4  & $1.09 $  &  -- & $ 1.56 $ & $ 1.39$ & $ 1.67 $  \\ \hline
			5  & $ 1.14$  & --  & $ 1.51  $ & $1.40 $ & $ 1.62 $  \\ \hline
			6  & $ 1.12$  &  --  & $ 1.50  $ & $1.48 $ & $ 1.58 $  \\ \hline
		\end{tabular}
	\end{center}
\end{table}

Table~\ref{hardmpt} presents hard-component $\bar p_{thi}(n_s)$ values derived from variable model functions $\hat H_{0i}(y_t,n_s)$, inferred from PID hard-component data in Sec.~\ref{tcmadapt}, integrated over 100 points distributed uniformly on $y_t \in [0,5]$ (i.e.\ within $p_t \in [0,10]$ GeV/c). Those values are used below in discussion of a PID \mmpt\ TCM.

\begin{table}[h]
	\caption{Spectrum hard-component means $\bar p_{thi}(n_s)$ for four hadron species from seven centralities of 5 GeV \ppb\ collisions with units GeV/c. $K^\pm$ values are copied from $K_\text{S}^0$ values because of the latter's much larger \yt\ coverage. These values are obtained by integrating over hard-component models from the variable TCM derived in this section.
	}
	\label{hardmpt}
	\begin{center}
		\begin{tabular}{|c|c|c|c|c|c|} \hline
			$n$	& $ \pi^\pm $  & 	$K^\pm$  & 	$p$  & 	$K_\text{S}^0$ & 	$\Lambda$  \\ \hline
			1  & $ 1.10$  & -- & $ 1.67 $ & $1.32 $ & $ 1.66 $  \\ \hline
			2  & $ 1.13$  &  -- & $ 1.65 $ & $ 1.33$ & $ 1.65 $  \\ \hline
			3  & $ 1.15 $  &  --  & $ 1.63 $ & $ 1.34$ & $ 1.63 $  \\ \hline
			4  & $1.18 $  &  -- & $ 1.61 $ & $ 1.35$ & $ 1.61 $  \\ \hline
			5  & $ 1.21$  & --  & $ 1.59  $ & $1.36 $ & $ 1.59 $  \\ \hline
			6  & $ 1.29$  &  --  & $ 1.55  $ & $1.39 $ & $ 1.56 $  \\ \hline
			7  & $1.30 $  & $ - $ & $ 1.51  $ & $1.42 $ & $ 1.55 $  \\ \hline
		\end{tabular}
	\end{center}
\end{table}

\begin{figure}[h]
	\includegraphics[width=1.65in,height=1.6in]{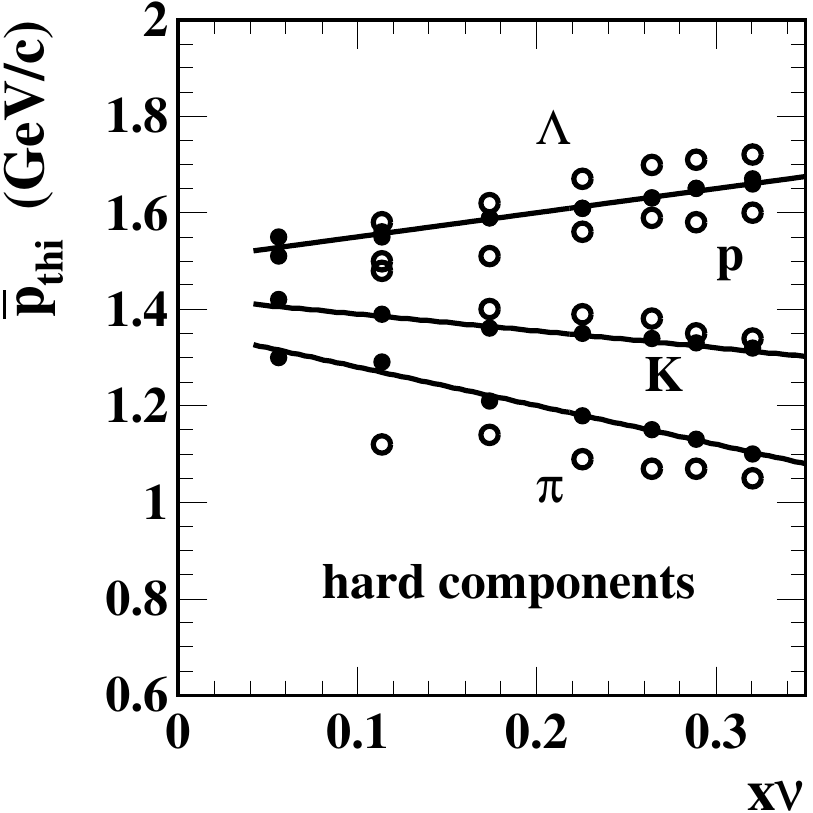}
	\includegraphics[width=1.65in,height=1.6in]{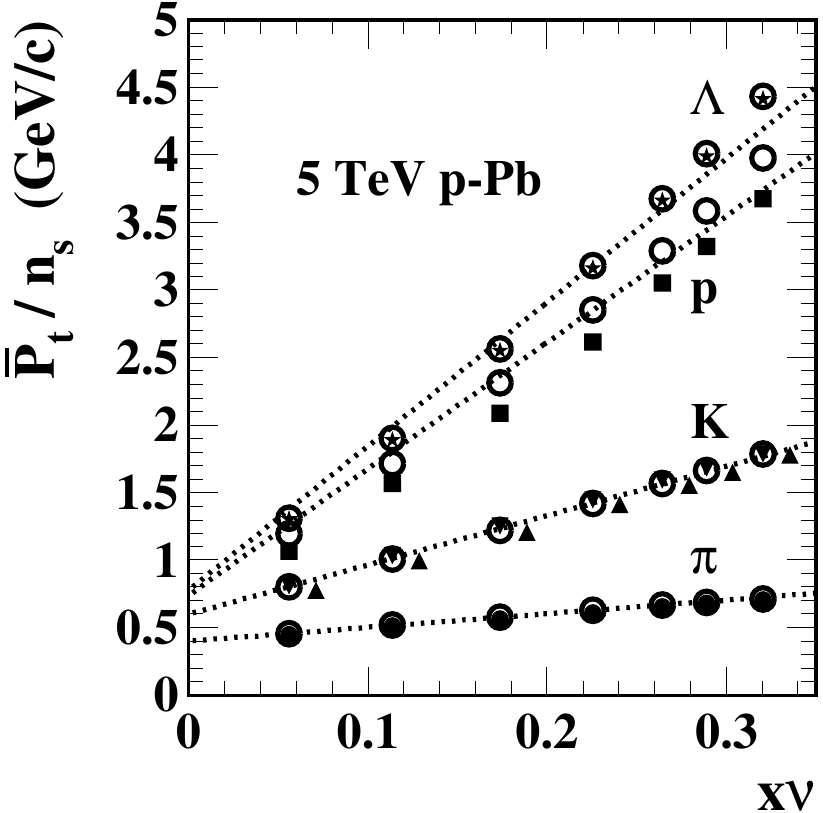}
	\caption{\label{qqq}
		Left: Mean-\pt\ hard components $\bar p_{thi}(n_s)$ for pions, kaons, protons and Lambdas from Tables~\ref{hardpt} (open circles) and \ref{hardmpt} (solid dots) vs hard/soft ratio $x\nu$. The lines are described in the text.
		Right:	Mean values $\bar P_{ti} / n_{si}$ as defined by Eq.~(\ref{pampttcmpid}) for four hadron species vs hard/soft ratio $x\nu$. The lines are described in the text.  The charged-kaon points are shifted to the right for visibility.
	} 
\end{figure}

Figure~\ref{qqq} (left) shows hard-component mean values $\bar p_{thi}(n_s)$ for four hadron species inferred from data presented in Table~\ref{hardpt} and from model functions derived in Sec.~\ref{tcmadapt} presented in Table~\ref{hardmpt}. Open points are obtained by simple integration of data hard components over their \pt\ acceptances and are therefore substantially biased. Solid points are obtained by integration``on a continuum'' of variable TCM hard components derived in Sec.~\ref{tcmadapt}. Lambda open points are high because the data lower bound is 0.65 GeV/c, and proton open points are low because of a systematic excess below the mode, both evident in Fig.~13 of Part I. Mean values generally increase substantially with hadron mass, and that trend is consistent with measured fragmentation functions for jets from \ee\ collisions at the large electron-positron collider LEP (see Fig.~7 of Ref.~\cite{ppbpid}).

Meson mean values decrease with centrality whereas baryon mean values increase. Those trends are a manifestation of hard-component centroid shifts reported in Sec.~\ref{tcmadapt} and are interpreted in Sec.~\ref{ratiotrends} to cause certain PID spectrum-ratio features. The trends (straight lines) for pions, kaons and protons/Lambdas respectively are $1.345 - 0.8 x\nu$, $1.425 - 0.35 x\nu$ and $1.50 + 0.5 x\nu$ GeV/c. 

\subsection{Total $\bf P_t$ vs soft hadron yield $\bf n_s$}

Especially in the context of small collision systems the relevance of an overarching ensemble mean $\bar p_{ti}(n_s)$ relating total $\bar P_{ti}$ to total charge $n_{chi}$ for hadron species $i$ can be questioned. The TCM for \ppb\ collisions describes a final state in which two components are closely related but clearly distinct. The structure of Eq.~(\ref{pampttcmid}) already relies on two distinct mean values -- $\bar p_{tsi}$ and $\bar p_{thi}$. A more-instructive mean-value definition that illustrates the simplicity of TCM structure has the form
\bea \label{pampttcmpid}
\frac{\bar P_{ti}}{n_{si}}
&=& [1 +  \tilde z_i(n_s)x(n_s)\, \nu(n_s)] \,  \bar p_{ti}(n_s).
\\ \nonumber
&=& \bar p_{tsi} + \tilde z_i(n_s) x(n_s)\nu(n_s) \, \bar p_{thi}(n_s),
\eea
where $n_{si} = z_{si} n_s$.
In what follows that relation is used to reexamine \pt\ mean-value trends for 5 TeV \ppb\ collisions.

Figure~\ref{qqq} (right) shows transformed data (points) and fixed TCM references (lines) evaluated according to Eq.~(\ref{pampttcmpid}). The solid points are $\bar p_{ti}(n_s)$ data from Fig.~4 of Ref.~\cite{aliceppbpid} converted to the format of Eq.~(\ref{pampttcmpid}) via its first line. The open points are derived from $\tilde z_i(n_s)$ values reported in Part I and from $\bar p_{thi}(n_s)$ values (solid points) from the left panel using $x(n_s)\nu(n_s)$ values from Table~\ref{rppbdata}. The lines are Eq.~(\ref{pampttcmpid}) (second line) evaluated with fixed parameter values $\tilde z_i(n_s)$ and $\bar p_{thi}(n_s)$ corresponding to centrality class $n = 4$. 
The baryon points fall below the fixed TCM reference for peripheral collisions and above the reference for central collisions. That trend is due to a combination of $\tilde z_i(n_s)$ and $\bar p_{thi}(n_s)$ (for baryons) increasing with centrality. In contrast, the meson points show no significant deviation from the fixed TCM reference.

\begin{figure}[h]
	\includegraphics[width=1.65in,height=1.58in]{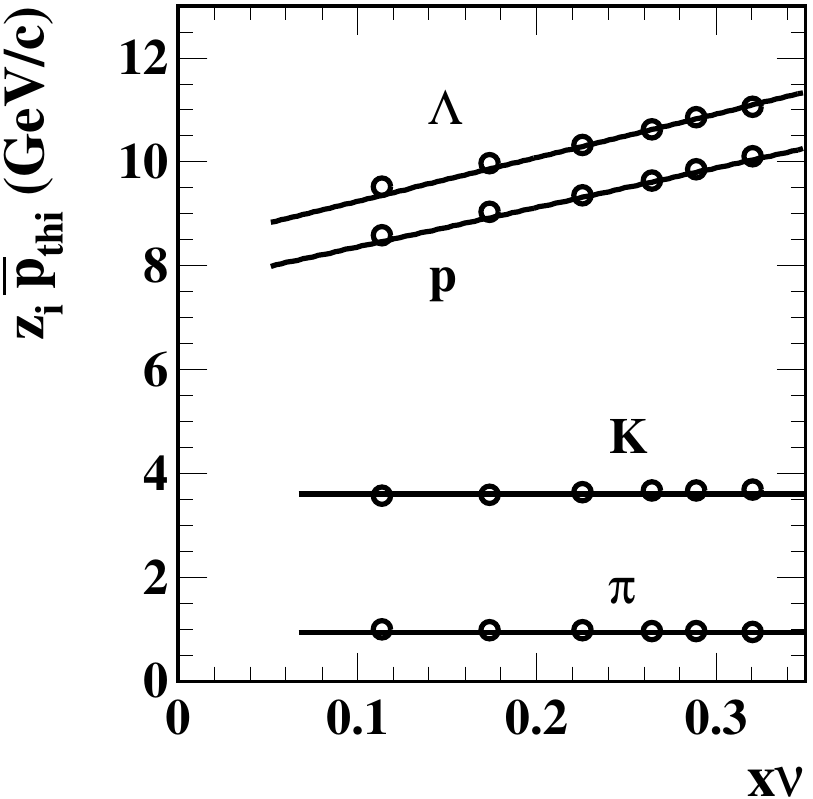}
	\includegraphics[width=1.65in,height=1.61in]{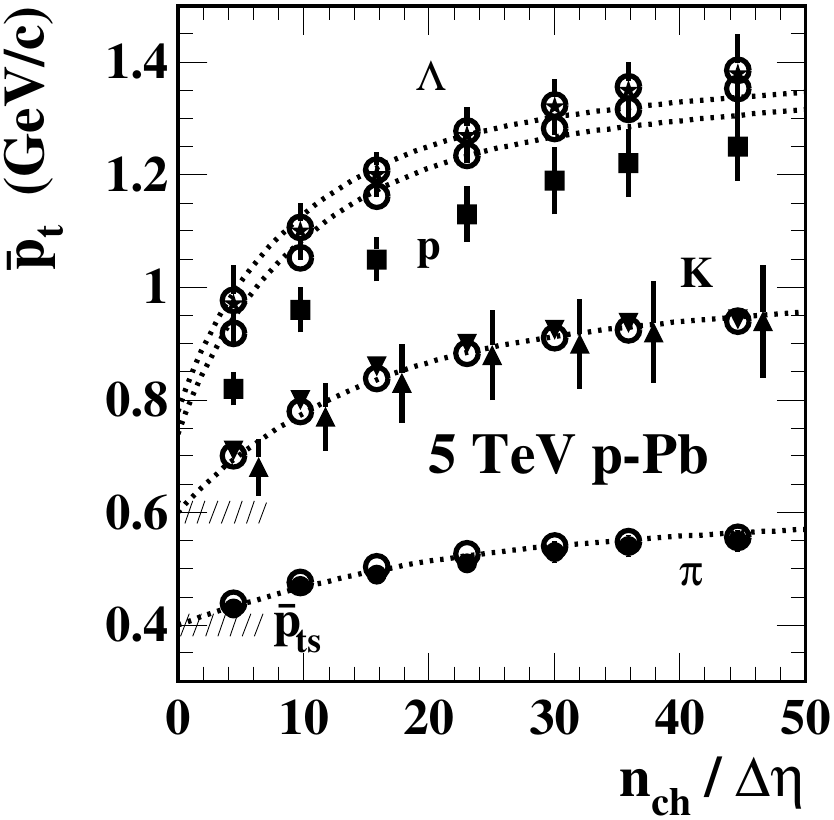}
	\caption{\label{rrr}
		Left: The product $\tilde z_{i}(n_s)\bar p_{thi}(n_s)$  for pions, kaons, protons and Lambdas with $\tilde z_{i}(n_s)$ from Fig.~8 of Part I and $\bar p_{thi}(n_s)$ from Table~\ref{hardmpt} vs hard/soft ratio $x\nu$. The lines are described in the text.
		Right:	Published $\bar p_{ti}$ data from Ref.~\cite{aliceppbpid} Fig.~4 (solid points)  and values inferred in the present study (open circles) from the same published PID spectra from Fig.~\ref{piddata} vs corrected charge density $\bar \rho_0 = n_{ch} / \Delta \eta$ for five hadron species. The curves are fixed TCM model Eq.~(\ref{pampttcmid}) with parameters corresponding to $n = 4$ and $\alpha = 0.0127$. The charged-kaon points are shifted to the right for visibility.
}  
\end{figure}

Figure~\ref{rrr} (left) provides the reason for the last observation. What determines the slope of the Eq.~(\ref{pampttcmpid}) $\bar P_{ti}/n_{si}$ trend on $x\nu$ is the {\em product} $\tilde z_i(n_s)\bar p_{thi}(n_s)$. Those trends are shown in Fig.~\ref{rrr} (left). Parameter $\tilde z_i(n_s)$ always increases in magnitude with centrality. When combined with baryon $\bar p_{thi}(n_s)$ increasing trends the baryon-product rate of increase is greater whereas combined with the meson $\bar p_{thi}(n_s)$ decreasing trends the meson-product variation is negligible. The straight-line trends for pions through Lambdas respectively are $0.93$, $3.6$, $7.6(1+x\nu)$ and $8.4(1+ x\nu)$ GeV/c. The various trends emphasize two issues: (a) jets (spectrum hard components) dominate \pt\ production in high-energy nuclear collisions, and (b) the nonPID hard/soft ratio $x\nu$ accurately describes the variation of jet production for four hadron species.

\subsection{Ensemble-mean $\bf \bar p_t$ TCM data description} \label{mmptdescription}

Figure~\ref{rrr} (right) shows published PID \mmpt\ data from Ref.~\cite{aliceppbpid} (solid points) and TCM results (open points and curves). The published error bars from Ref.~\cite{aliceppbpid} are included in this format. The TCM points and lines in Figure~\ref{qqq} (right) are transformed to conventional ensemble mean $\bar p_{ti}(n_s)$ via Eq.~(\ref{pampttcmpid}) (first line). Comparison of the present study with results of Ref.~\cite{aliceppbpid} is as follows: The kaon result from the present study (open circles) is statistically indistinguishable from the published $K_\text{S}^0$ data (inverted solid triangles). Both trends are slightly above the TCM reference for peripheral and slightly below the reference for central collisions. While the product $\tilde z_i(n_s)\bar p_{thi}(n_s)$ for mesons has negligible variation (left panel) conventional mean $\bar p_{ti}(n_s)$ includes the expression $1 +  \tilde z_i(n_s)x(n_s)\, \nu(n_s)$ in its denominator, the increasing $\tilde z_i(n_s)$ leading to an overall decrease for kaons and pions.

The published Lambda data agree with the present study within their systematic uncertainties as do the charged-kaon data. However, uncorrected proton data fall well below the present results, presumably due to the detection inefficiency identified in Part I.  The baryon data rise substantially faster than the TCM reference as expected from the left panel. The published pion data agree well with the present study given the effective soft-component value $\bar p_{tsi} \approx 0.40$ GeV/c. In Part I Sec.~III it was demonstrated that pion spectra require an explicit model for a resonance contribution. Without that contribution $\bar p_{tsi} \approx 0.44$ would clearly disagree with \mmpt\ data.

\section{Systematic uncertainties} \label{sys}

The present study (Part II) considers systematic uncertainties throughout the text. Three issues are considered in this section: (a) The TCM parameter inventory, its origins and uncertainties; (b) TCM prediction accuracy and determination of data ``fine structure'' and (c) TCM physical interpretations vs other spectrum models.

\subsection{TCM parameter inventory}

Different elements of the TCM may arise from different collision systems that are subsequently unified within an overall physical context. The full spectrum TCM employed in the current study is defined in Eq.~(\ref{pidspectcm}) (second line). Its several elements are discussed below.

NonPID charge densities $\bar \rho_s$ and $\bar \rho_h$ (and hence \ppb\ ``geometry'' in terms of jet production) are based on 5 TeV \ppb\ \mmpt\ data analyzed in Ref.~\cite{tommpt}. The \ppb\ \mmpt\ data are described there by the TCM within data uncertainties out to $\bar \rho_0 \approx 115$ (in contrast to data extending only to $\bar \rho_0 \approx 45$ for Ref.~\cite{aliceppbpid}). Indicators of the quality of the \ppb\ geometry description are the low-\yt\ tails of the $K^0_\text{S}$ hard components in Fig.~\ref{newkaon} (left) and the high-\yt\ tails of the Lambda hard components in Fig.~\ref{newprot} (left). The hard components are only rescaled by factors $\bar \rho_h$ and $\hat H_{0i}(\bar y_t)$. The latter are peak amplitudes all near 0.29 that vary by a few percent and only with variation of the widths of the $\hat H_{0i}(y_t)$ model functions. Thus, the stability of the PID hard-component tail structure in the two cases is indicative of the accuracy of the nonPID $\bar \rho_h$ estimates.

Coefficients $z_{si}(n_s)$ and $z_{hi}(n_s)$ are \nch-dependent PID {\em fractions} of soft and hard charge densities $\bar \rho_s$ and $\bar \rho_h$. The analysis in Part I demonstrates that those coefficients are accurately generated by simple functions of hard/soft ratio $x\nu$ with linear dependence on hadron mass $m_i$, determined by only two parameters, and on hadron species fractions $z_{0i}$ comparable to statistical-model predictions. Those coefficients were demonstrated to be self-consistent at the percent level in Part I (e.g.\ its Fig.~10, right).

Fixed TCM model functions $\hat S_{0i}(y_t)$ and $\hat H_{0i}(y_t)$ are determined by comparisons with data. However, they are not fitted to individual spectra or collision systems. In Table II of Part I the model parameters for nonPID hadrons ($h$) are obtained from Ref.~\cite{alicetomspec}, a systematic survey of \pt\ spectra from \pp\ collisions over a range of collision energies. The {\em interpolated} nonPID values for 5 TeV \pp\ collisions (not directly measured) are compatible with the PID values for 5 TeV \ppb\ collisions reported in Table~\ref{pidparamsxx} of the present study. The TCM, serving as a {\em fixed reference}, is thus compatible with an array of collision systems and collision energies within data uncertainties.

In the present study PID TCM hard components are varied systematically to accommodate data as reported in Sec.~\ref{tcmadapt}. The resulting parameter variations are simply described in terms of linear variation with hard/soft ratio $x\nu$ as in Fig.~\ref{widvar}. Once again, the TCM does not rely on fits to individual spectra for its definition. The accuracy of the description is made clear by Z-score distributions reported in Sec.~\ref{modelacc}.

\subsection{TCM prediction accuracy}

The variable TCM adjusted to accommodate PID spectrum data can be used to predict secondary data features, one example being ensemble \mmpt\ trends as in Sec.~\ref{pidmmpt}. Hard-component means $\bar p_{thi}(n_s)$ derived from variable TCM model functions in Sec.~\ref{tcmadapt} are combined in Eq.~(\ref{pampttcmpid}) with nonPID $x\nu$ presented in Table~\ref{rppbdata} and $\tilde z_i(n_s)$ trends from Part I to produce PID \mmpt\ trends corresponding to spectra extrapolated over $y_t \in [0,5]$. In Fig.~\ref{qqq} (right) the resulting entries (open points) are compared with a fixed TCM reference (dotted lines) to obtain ``fine structure'' of the \mmpt\ trends. Systematic deviations from the fixed TCM are explained by the linear (on $x\nu$) trends in Fig.~\ref{rrr} (left) to reveal \mmpt\ features at the percent level.

Another example is spectrum ratio trends as in Sec.~\ref{specratios}. The TCM ratio model defined in Eq.~(\ref{ratiopidtcm}) is used to reproduce the centrality asymmetry that appears in Fig.~\ref{ratios} and to trace its origins to translation of hard-component peak centroids to lower or higher \yt\ with increasing \nch\ that is revealed in Sec.~\ref{tcmadapt}. As with \mmpt\ data, sensitivity at the statistical limits of spectrum data is possible via differential studies relative to the fixed TCM as reference.

\subsection{TCM physical interpretation}

An important aspect of systematic uncertainties is the {\em interpretability} of model results. One may ask whether the model has a simple structure and well-defined implementation so that different groups may obtain equivalent results with the same model applied to equivalent data. An example of substantial disagreement is blast-wave (BW) model analysis of PID spectra from 13 TeV \pp\ collisions in Refs.~\cite{cleymans} and \cite{alicepppid}. Equation~(14) of Ref.~\cite{cleymans} defining the BW model is markedly different in form from Eq.~(1) of Ref.~\cite{aliceppbpid} repeated from Ref.~\cite{blastwave}. As noted in Fig.~12 of Ref.~\cite{tommodeltests} BW model parameters $\langle \beta_t \rangle$ and $T_{kin}$ inferred in two cases differ by a factor 2 or more, providing one reason to question the model's validity.

In contrast, the TCM as defined in this and previous studies has a simple and clear algebraic structure that makes possible differential extraction of all information contained in particle data down to statistical uncertainties, as demonstrated in Sec.~\ref{modelacc}. The TCM soft component is arguably related to longitudinal dissociation of projectile nucleons, a manifestation of an eventwise PDF in terms of a soft hadron distribution. Its \pt\ spectrum (i.e.\ L\'evy distribution) is consistent with thermal emission from an incompletely thermalized source or a source structured due to a longitudinal parton splitting cascade. Note that for exponent $1/n \rightarrow 0$ the L\'evy distribution goes to a Maxwell-Boltzmann distribution on \mt. The TCM hard component is quantitatively comparable with a convolution of {\em measured} fragmentation functions with a {\em measured} parton/jet energy spectrum~\cite{fragevo}. The direct quantitative connection between spectrum structure at low \pt\ (e.g.\ $\approx 0.2$ GeV/c) and the jet contribution at higher \pt\ (e.g.\ above 4 GeV/c), as described in Part I, Secs.~II D, III B and IV B, seems to preclude any role for hydrodynamic expansion in \ppb\ collisions.

\section{Discussion}  \label{disc}

This section compares text in Ref.~\cite{aliceppbpid} relating to interpretation of PID spectrum trends with some alternative viewpoints. Relevant issues include full utilization of information carried by PID particle data and acknowledgement of substantial jet contributions to nuclear collisions as determined by earlier high-energy physics (HEP)  research and as described by conventional QCD principles.

\subsection{$\bf p$-Pb $\bf p_t$ spectra and the QGP narrative} \label{narrative}

Reference~\cite{aliceppbpid} presents a certain narrative relating to QGP formation in high-energy nuclear collisions in discussing its \ppb\ PID spectrum measurements. Quoted text below is labeled A, I, R, D, C for paper Abstract, Introduction, Results, Discussion and Conclusions.

The importance of \pt\ spectra for interpreting high-energy collision data is acknowledged: (I) ``The $p_T$ distributions and yields of particles of different mass at low and intermediate momenta of less than a few GeV/c (where the vast majority of particles [e.g.\ jet fragments] is produced), can provide important information about the system created in high-energy hadron reactions.'' Also acknowledged is the importance of smaller systems for interpreting results from heavy-ion collisions: (I) ``The interpretation of heavy-ion results depends crucially on the comparison with results from smaller collision systems such as proton-proton (pp) or proton-nucleus (pA).'' However, the effectiveness of small systems for testing claims associated with A-A collisions is effectively compromised: (I) [because the charge density in \pa\ is high] ``Therefore the assumption that final state dense matter effects can be neglected in pA may no longer be valid.''

A broader context is established with presentation of the QGP scenario believed by many to describe at least heavy-ion collisions: (I) ``The bulk matter created in high-energy nuclear reactions can be quantitatively described in terms of hydrodynamic and statistical models. The initial hot and dense partonic matter rapidly expands and cools down, ultimately undergoing a transition to a hadron gas phase.'' Presumed rapid expansion of hot and dense partonic matter is then related to spectrum structure: (I) ``This [collective hydrodynamic flow] results in a characteristic dependence of the shape of the transverse momentum ($p_T$) distribution on the particle mass....'' (A) ``The transverse momentum distributions exhibit a hardening as a function of event multiplicity, which is stronger for heavier particles. This behavior [in \ppb\ collisions] is similar to what has been observed in pp and Pb-Pb collisions at the LHC.'' (R) The [\ppb] $p_T$ distributions show a clear evolution, becoming harder as the multiplicity increases. The change is most pronounced for protons and lambdas. They show an increase of the slope [actually {\em decrease} of the slope] at low $p_T$, similar to the one observed in heavy-ion collisions.''
(R) ``The latter [in \pbpb\ collisions]  are generally discussed  in terms of collective flow or quark recombination.''

The possibility of hydrodynamic expansion affecting spectrum features then leads to introduction of the blast-wave spectrum model: (D) ``In heavy-ion collisions, the flattening [i.e.\ {\em decreased} slope] of transverse momentum distribution and its mass ordering find their natural explanation in the collective radial expansion of the system. This picture can be tested in a blast-wave framework with a simultaneous fit to all particles for each multiplicity bin. This parametrization assumes a locally thermalized medium, expanding collectively with a common velocity field and undergoing an instantaneous common freeze-out.'' (R) ``Several parametrizations [of spectra] have been tested, among which the blast-wave function...gives the {\em best description of the data over the full $p_T$ range}'' [emphasis added, fits are independently applied to individual spectra]. (D) ``In contrast with the individual [blast-wave] fits discussed above, the simultaneous fit to all particle species under consideration can provide insight on the (common) kinetic freeze-out properties of the system.'' ``...the actual values of the fit parameters depend substantially on the fit range [on \pt].'' (D) ``As can be seen in Fig.~6, the parameters show a similar trend as the ones obtained in Pb-Pb. Within the limitations of the blast-wave model, this observation is consistent with the presence of radial flow in p-Pb collisions.'' However, as noted in Ref.~\cite{hydro}, in the presence of radial flow ``...the particle yield is thus depleted at low $p_T$. The heavier the particle and the larger the flow velocity $v_T$, the larger the depletion.'' Thus, precise measurements of PID spectrum structure down to very low \pt\ {\em compared to a reference} would be essential to claim radial flow in \ppb\ collisions. However, the fact that for example proton $z_{si(n_s)}$ values obtained near 0.15 GeV/c lead to accurate predictions of jet fragment yields near 4 GeV/c (Part I, Sec.~III B) makes radial flow in \ppb\ collisions quite unlikely.

Within quoted text above one encounters a tendency to emphasize similarities between data trends in small and large systems with the implication that QGP formation may then occur as well in the former. But when such comparisons lead to counterintuitive results {\em ad hoc} explanations are introduced: (D) Comparing \ppb\ and \pbpb\ data ``...at similar $dN_{ch}/d\eta$ the [blast-wave] $\langle \beta_T \rangle$ values are significantly higher in \ppb\ collisions. While in \pbpb\ collisions high multiplicity events are obtained through multiple soft interactions, in \ppb\ collisions the high multiplicity selection biases the sample towards harder [\pn] collisions. This could lead to the larger $\langle \beta_T \rangle$ parameter obtained from the blast-wave fits. Under the assumptions of a collective hydrodynamic expansion, a larger radial velocity in \ppb\ collisions has been suggested as a consequence of stronger radial gradients.''

The conclusions of Ref.~\cite{aliceppbpid} include (C) ``These data represent a crucial set of constraints for the modeling of proton-lead collisions at the LHC. The transverse momentum distributions show a clear evolution with multiplicity, similar to the pattern observed in high-energy pp and heavy-ion collisions, where in the latter case the effect is usually attributed to collective radial expansion.''

\subsection{Full utilization of spectrum information}

Reference~\cite{aliceppbpid} provides little significant {\em quantitative} information actually derived from spectrum data. Spectra are described qualitatively, and in multiple instances, as ``hardening'' with increasing \nch, increasingly so with increasing hadron mass. Spectrum data are said to provide a  ``crucial set of constraints'' but no specifics are provided. The word ``constraints'' could be associated with information extracted from spectrum data, but corresponding information is not made apparent. The article concludes with the statement ``The transverse momentum distributions show a clear evolution with multiplicity...,'' but nothing clear emerges from the paper's presentation about spectrum evolution {\em or its causes}.

In contrast, material presented in Part I (Part I) and Secs.~\ref{tcmadapt}, \ref{specratios} and \ref{pidmmpt} of the present study (Part II) provides exhaustive detail on spectrum structure and its evolution with \nch. Spectrum hard components, previously related quantitatively to measured jet properties~\cite{eeprd,hardspec,fragevo}, are isolated accurately at the level of spectrum-data statistical uncertainties. That \ppb\ \pt\ spectra ``harden'' with increasing \nch\ is a simple consequence of the quadratic relation between the nonjet (soft) hadron yield and jet production~\cite{ppprd}, which in turn is reflected in the measured systematics of jet cross sections for high-energy \pp\ collisions~\cite{jetspec2}. That the ``hardening'' trend increases with hadron mass again corresponds to hard-component (i.e.\ jet fragmentation) systematics as reflected in the $\tilde z_i(n_s)$ trends presented in Part I Sec.~V (e.g.\ its Fig.~8).

Substantial effort is extended in Ref.~\cite{aliceppbpid} to extract information {\em of a sort} in the form of spectrum and yield ratios (its Figs.~2, 3 and 5). However, the physical mechanisms that determine such ratio evolution remain obscure within that presentation. Referring to its Fig.~2, Ref.~~\cite{aliceppbpid} asserts that ``The ratios $p/\pi$ and $\Lambda/K^0_\text{S}$ show a significant enhancement [relative to what?] at intermediate $p_T \sim 3$ GeV/c, {\em qualitatively reminiscent} of that measured in \pbpb\ collisions [emphasis added]. The latter are generally discussed in terms of collective flow or quark recombination.'' While the features are indeed {\em qualitatively} similar for two collision systems the underlying mechanism(s) are not {\em substantively} identified. Similarly, a ``power-law'' trend is inferred from \nch\ dependence of the \ppb\ $p/\pi$ spectrum ratio as in Fig.~3 of Ref.~\cite{aliceppbpid} and compared with a similar \pbpb\ trend. But the origin of the trend in either system is not determined.

In contrast, Sec.~\ref{specratios} provides a detailed explanation of spectrum ratio shapes and evolution with \nch\ in terms of PID spectrum hard components. The well-known $p/\pi$ peak near 3 GeV/c is a simple consequence of the separation of corresponding pion and proton hard-component modes at $y_t \approx 2.46$ and 3.0 ($p_t \approx 0.80$ and 1.4 GeV/c) respectively combined with the difference in their power-law tails, as demonstrated in Sec.~\ref{ratiotrends}. Centrality dependence of the $p/\pi$ spectrum ratio reflects shifts to higher and lower \yt\  respectively of the proton and pion hard components that are clearly demonstrated in Sec.~\ref{tcmadapt}. The power-law trend in Fig.~3 of Ref.~\cite{aliceppbpid} is a consequence of the Gaussian shape of spectrum hard components as demonstrated in Fig.~\ref{xxx} of the present study. That there are similarities between trends in \ppb\ data and in \pbpb\ data suggests that \pbpb\ spectra be analyzed in similar ways to extract {\em jet-related hard components} for direct comparison. But that has already been done in Ref.~\cite{pbpbpid}.

Various ratio formats invoked by Ref.~\cite{aliceppbpid} may respond to interesting spectrum features relating to \ppb\ centrality dependence but also tend to discard essential information carried by spectrum data.  Spectrum ratios alone may not be interpretable within the conventional context. In contrast, differential TCM analysis makes clear (at the percent level) what aspects of collision dynamics produce those manifestations and quantitatively how those features are related to fundamental QCD processes.

\subsection{Jet contributions to nuclear collisions}

The narrative introduced in Ref.~\cite{aliceppbpid} and summarized in Sec.~\ref{narrative} invokes {\em argument from analogy} to interpret \ppb\ spectrum data in terms of a popular interpretation of \aa\ data as demonstrating QGP formation (hot and dense partonic matter) and transverse flows. Such arguments reverse a convention adopted upon commencement of the relativistic heavy ion collider (RHIC) program that inference of QGP formation in \aa\ collisions (e.g.\ \auau\ collisions) based on certain data features should depend on {\em absence} of those features in smaller collision systems (e.g.\ \dau\ collisions~\cite{daufinalstate}) serving as controls. With the appearance of certain features in \ppb\ and even \pp\ collisions at the LHC the control concept has been abandoned and interpretations based on QGP formation are favored. In retrospect, there was lacking a full understanding of the jet physics of small collision systems, at least among those pursuing research within the RHIC program.

The larger context is a conflict between two narratives -- the popular flow/QGP narrative and a QCD-based jet-production narrative -- that compete to explain certain critical data features. Arguments such as those summarized in Sec.~\ref{narrative} tend to favor the flow/QGP narrative as a desired outcome and provide no significant role for minimum-bias jet formation. For instance, it is argued that [because the ``charge density'' in \pa\ is high] ``Therefore the assumption that final state dense matter effects [i.e.\ QGP formation] can be neglected in pA may no longer be valid.'' But as noted above the increasing $\bar \rho_0$ in \ppb\ collisions is due primarily to the soft component, which shows no sign of modification in any system, and secondarily to the jet-related hard component whose minor evolution with \nch\ (see Sec.~\ref{tcmadapt}) can be interpreted in terms of well-known jet properties. The presence of the Pb nucleus {\em per se} is apparently irrelevant. What jet (hard-component) evolution {\em is} evident varies smoothly from charge densities {\em below NSD} to a ten-fold increase.

The signature manifestation of QCD in high-energy hadron-hadron collisions is jet production. The physics of jet production in hadronic collisions has been established quantitatively over several decades by the HEP community. If small systems are to serve as meaningful controls jet production within such systems must be fully understood by those embarking on a search for QGP formation in heavy-ion collisions.

Transport of longitudinal projectile momentum to transverse degrees of freedom is a fundamental issue. The QGP scenario invokes particle (hadron or parton?) energy loss (``stopping'') during the collision followed by partial equilibration to a high-density medium and consequent generation of transverse matter/energy gradients that drive radial expansion prior to final-state hadron formation. That scenario, if valid, might result in certain modifications of hadron spectra invoked as signals of QGP formation, for instance spectrum ``hardening'' and mass dependence as described in Ref.~\cite{aliceppbpid}.

In contrast, minimum-bias dijet production provides a mechanism for energy transport from longitudinal to transverse degrees of freedom in elementary collisions that is {\em predicted} by conventional QCD theory and has been thoroughly investigated via HEP jet measurements. MB dijet production produces PID spectrum manifestations somewhat similar to what is expected for QGP formation: hardening of spectra with increasing \nch\ and with increasing hadron mass. But those trends are {\em quantitatively predicted} by the quadratic relation between jet production in elementary collisions and nonjet soft hadron production (projectile dissociation -- longitudinal hadronization of low-$x$ gluons) -- as in the TCM.

The discussion in Ref.~\cite{aliceppbpid} includes the assertion ``In heavy-ion collisions, the flattening of transverse momentum distribution and its mass ordering find their {\em natural explanation} in the collective radial expansion of the system [emphasis added].'' But in elementary collisions (at least) similar spectrum trends ``find their natural explanation'' in QCD-predicted dijet production as demonstrated by the present study and others~\cite{fragevo,jetspec2,mbdijets}. As noted, the comparison between small and large collision systems represents, in effect, a competition between MB diject production and QGP formation to explain pivotal data manifestations. If \ppb\ data look like \pbpb\ data {\em in some sense} does that mean \ppb\ collisions exhibit physics {\em assumed} for \pbpb\ collisions (what is implied in Ref.~\cite{aliceppbpid}) or do \pbpb\ collisions actually manifest a variant of simple QCD physics {\em predicted} for elementary collisions? 

To buttress claimed QGP formation in \aa\ collisions one would have to master minimum-bias jet contributions to small collision systems and then demonstrate that something truly novel occurs in larger collision systems. The ambiguity between transport mechanisms and their manifestations described above may be resolved by sufficiently differential analysis, as in the present study, applied consistently to small and large collision systems.

\subsection{Baryons, strangeness and jets}

The centrality trends of coefficients $z_{si}(n_s)$, $z_{hi}(n_s)$ and hard/soft ratio $\tilde z_i(n_s)$ in Part I Sec.~V A reveal the dominant role of jet production in the abundance of baryons and strangeness in high-energy nuclear collisions. PID hard/soft ratio $\bar \rho_{hi} / \bar \rho_{si} = \tilde z_i(n_s) x \nu$ is the key parameter. \ppb\ nonPID ratio $x\nu = \bar \rho_{h} / \bar \rho_{s}$ ranges from about 0.06 to 0.32 for all hadron species. $\tilde z_i(n_s)$ however increases with hadron mass from about 0.8 for pions to 6 for baryons. There is apparently no sensitivity to strangeness {\em per se}. The $z_{xi}(n_s)$ trends are determined solely by hadron mass $m_i$ and overall abundances $z_{0i}$.

For most-central \ppb\ collisions (at least for the spectrum data reported in Ref.~\cite{aliceppbpid}) and $x\nu \approx 0.32$ the hard/soft (jet/nonjet) ratio for Lambdas (for instance) is $6.5 \times 0.32 \approx 2.1$ for {\em integrated yields}. Above 1 GeV/c Lambda spectra are completely dominated by jet fragments. For kaons and central collisions the yield ratio is $2.5 \times 0.32 \approx 0.8$, and again spectra above about 1 GeV/c are dominated by jet fragments. Even for pions the jet contribution plays a dominant role above 2 GeV/c.

Minimum-bias jets appearing near midrapidity in high-energy \nn\ collisions have their origin in low-$x$ gluons as part of eventwise PDFs. The results in Part I Sec.~V A (among others) indicate that large-angle scattered gluons in the form of MB dijets produce large amounts of \pt, strangeness and baryons near midrapidity. Within the flow/QGP narrative those same indicators may be associated not with jets but with spectrum ``hardening'' (due to radial flow), parton recombination and strangeness enhancement in a dense QCD medium or QGP. To resolve that conflict requires detailed differential spectrum analysis as reported in the present study and Part I.

\section{Summary}\label{summ}

This article presents Part II of a two-part study of identified-hadron (PID) \pt\ spectra from 5 TeV \ppb\ collisions.  In Part I a previously-reported preliminary PID spectrum two-component (soft + hard) model (TCM) was extended in several ways, including incorporation of a resonance model for pions, correction of proton spectra for detection inefficiency and determination of TCM coefficients $z_{si}(n_s)$ and $z_{hi}(n_s)$ directly from spectra. Minimum-bias data hard components $H_i(y_t,n_s)$ were extracted from spectrum data with an improved method, and their evolution with \ppb\ centrality was determined relative to a fixed PID TCM serving as a reference. 

Whereas the analysis described in Part I is based on the TCM as a fixed model, with data deviations suggesting novel physics,  the analysis described in this article (Part II) includes adjustment of the TCM to describe all data accurately within their statistical uncertainties as determined by standard statistical measures (Z-scores). No physically-significant data-model deviations remain.

Based on the differential study in Part I it is evident that different model parametrizations are required for meson and baryon hard components. However, the models for strange and nonstrange baryons are identical. The small-but-significant model differences between strange and nonstrange mesons may be entirely due to the substantial mass difference between pions and kaons. The parameter variations can be expressed as simple linear functions of TCM hard/soft (jet/nonjet) ratio $x\nu$.

Given its precise representation of PID spectrum data for \ppb\ collisions, including not only actual data acceptances but accurate extrapolations over the interval $y_t \in [0,5]$ ($p_t \in [0,10]$ GeV/c), the variable TCM is then applied to a study of PID spectrum and yield ratios and ensemble-mean \mmpt\ trends. Variation of \pt\ spectrum ratios with charge multiplicity \nch\ is predicted by the TCM. Certain notable features of PID spectrum ratios are thereby explained, including the asymmetry in centrality evolution centered (for proton/pion) near \yt\ = 3.1 ($p_t \approx 1.55$ GeV/c) and the prominent peak near 3 GeV/c. The former arises from shifts of jet-related hard components with increasing \nch\ to higher \yt\ for baryons in the numerator vs shifts to lower \yt\ for mesons in the denominator. The approximate power-law dependence on \nch\ of spectrum ratios for specific \pt\ values is simply due to the approximate Gaussian shape of spectrum hard components. A prominent peak near 3 GeV/c arises because of two factors: pion and proton hard-component peak modes are well separated on \yt, and the pion peak width is substantially greater than the proton width. The Lambda-to-kaon ratio features are similarly explained.

PID yield ratios are likewise predicted by the TCM in terms of fractions $z_{0i}$ that should be directly comparable with the statistical model of hadron production. A notable consequence of the TCM description is that total-yield ratios should be independent of \nch\ or \ppb\ centrality. Significant deviations from that prediction suggest possible biases in the processing of particle data.

PID ensemble-mean \mmpt\ trends are similarly predicted by the TCM. Comparisons of PID \mmpt\ data to a fixed TCM reference reveal subtle details that are nevertheless fully explained by variation with \nch\ of hard/soft ratios $\tilde z_i(n_s) \equiv z_{hi}(n_s) / z_{si}(n_s)$ as well as hard-component peak centroids and shapes as established in the present study.

Several conclusions can be drawn from the combined study reported in Parts I and II: 

(a) Variations of strangeness production (enhancement and suppression) and baryon abundance are conventionally associated with possible quark-gluon plasma (QGP) formation. The present study reveals that strong variation of strangeness and baryon number with \ppb\ collision \nch\ is dominated by jet production trends, and jet-related strangeness and baryon production in turn depends only on hadron mass. Thus, strangeness {\em per se} and baryon identity (not {\em net} baryon number) {\em per se} do not significantly determine hadron abundance in smaller collision systems. Hadron mass is the determining factor.

(b) The only ``centrality'' (i.e.\ \nch) dependent aspect of spectrum structure relative to a fixed PID TCM reference is evolution of hard-component shapes, differently for meson and baryon hard components. The basics of jet formation in the context of a QCD description of scattered-parton production and fragmentation suggest that the shape evolution arises mainly from control of event-wise parton distribution functions (PDF) by the choice of average event charge multiplicity \nch.

(c) Evolution of certain \pt\ spectrum features with \mbox{A-B} centrality (e.g.\ increased spectrum ``hardening'' -- to a greater extent for more-massive hadrons) has been attributed to radial flow. The present study reports a TCM based only on longitudinal projectile-nucleon dissociation and transverse jet production that provides exhaustive descriptions of four hadron species for seven centrality classes of 5 TeV \ppb\ collisions. The TCM is thus {\em sufficient} as a spectrum model. An earlier study of \pp\ spectrum data not based on {\em a priori} assumptions demonstrated that the TCM emerges as a {\em necessary} model. No other model provides equivalent accuracy and universality. Since all information carried by spectrum data is fully represented by the TCM one may conclude that the PID \ppb\ spectrum data addressed in the present study provide no significant evidence for radial flow.


\end{document}